\documentclass[aps,prb,twocolumn,groupedaddress]{revtex4}
\usepackage[dvips]{graphicx} \usepackage{epsfig}

\begin{document}

\title{Correlations between the morphology and the electronic structure at the surface
of thin film manganites, investigated with STM.}

\author{S. Kelly, F. Galli, I. Komissarov, J. Aarts}

\affiliation{Department of Physics, Leiden University, \\ 2300 RA Leiden,
Netherlands}

\date{\today}

%%%%%%%%%%%%%%%%%%%%%%%%%%%%%%%%%%%%%%%%%%%%%%%%%%%%%%%%%%%%%%%%%%%%%%%%%%%%%%%%%%%%%%
\begin{abstract}
%%%%%%%%%%%%%%%%%%%%%%%%%%%%%%%%%%%%%%%%%%%%%%%%%%%%%%%%%%%%%%%%%%%%%%%%%%%%%%%%%%%%%%

Thin-film colossal magnetoresistance manganites such as
La$_{0.67}$Ca$_{0.33}$MnO$_{3}$ (LCMO) have now been intensely studied for more than
a decade, but the issue of possible nanoscale electronic phase separation is not
fully solved. Scanning Tunneling Microscopy / Spectroscopy (STS) has been pivotal in
studying phase separation, but is hindered by being surface- rather than
bulk-sensitive. For our sputtered LCMO films the data indicates a strong correlation
between surface morphology and signatures of phase separation; rough films show
phase separation while atomically flat films are electronically homogeneous but have
a more or less inactive surface layer. Regardless of surface morphology, the film
bulk is electronically and magnetically active. Many of the reported conclusions
about electronic inhomogeneities measured by STS have been confused by this issue.
We study both strained and unstrained films and find no correlation between
substrate-induced strain and either electronic phase separation or dead layers.

 \end{abstract}

%\keywords{}

\maketitle

%%%%%%%%%%%%%%%%%%%%%%%%%%%%%%%%%%%%%%%%%%%%%%%%%%%%%%%%%%%%%%%%%%%%%%%%%%%%%%%%%%%%%%
\section{Introduction}
%%%%%%%%%%%%%%%%%%%%%%%%%%%%%%%%%%%%%%%%%%%%%%%%%%%%%%%%%%%%%%%%%%%%%%%%%%%%%%%%%%%%%%
Hole-doped manganites, such as La$_{0.67}$Ca$_{0.33}$MnO$_{3}$, have been widely
studied, not only because they exhibit colossal magnetoresistance (CMR), but also
because of interest in the coupled metal-insulator ($M$-$I$) and
ferromagnetic-paramagnetic transition they demonstrate. One unanswered question is
how electronic phase separation (PS) occurs, that is its spatial structure as
function of temperature and magnetic field, and how that connects to the CMR effect
observed in transport measurements. Local-probe techniques, such as scanning
tunneling microscopy (STM) and magnetic force microscopy, would seem to be ideal
tools to explore this, and a number of groups have studied thin-film manganites
using these
methods.\cite{Fath1999,Becker2002,Chen2003,Seiro2008,Biswas2001,Zhang2002} Some
reports find PS on the scale of many nanometers up to micrometers. Because of the
electrostatic energy cost of domains on the order of micrometers, phase separation
should be limited to nanometers, unless large disorder is
present,\cite{Moreo2000,Dagotto2008} suggesting that disorder must play a role in
such large-scale PS. However, extensive investigations, both experimentally and
theoretically, also indicate that manganite surface properties can differ from bulk
properties.\cite{Borges2001,Abad2005,Yao2006,Valencia2007a,Choi1999,Simon2004,Estrade2007,Calderon1999,Filippetti2000,Zenia2005}
This is of obvious relevance, since techniques such as scanning tunneling microscopy
(STM) and spectroscopy (STS) are inherently surface-sensitive. Nevertheless, from
STM and STS measurements of thin-film manganites, evidence of
half-metallicity,\cite{Wei1997} polarons,\cite{Wei1997,Seiro2008} and
pseudo-gaps,\cite{Mitra2005} have all been claimed, while spatial electronic phase
separation has been mapped as a function of both applied magnetic field,
\cite{Fath1999,Chen2003} and temperature.\cite{Becker2002} These surface
measurements are purported to represent bulk properties, but we want to reconsider
this issue, which has been further clouded because STM/STS measurements have been
made on a variety of systems. It should be realized that spatial inhomogeneities can
be significantly different for wide bandwidth systems such as La$_{1-x}$Sr$_{x}$MnO
or narrow bandwidth systems such as La$_{1-x}$Ca$_{x}$MnO$_{3}$; or for differing
strain states such as fully strained versus (partially) relaxed. We present an
STM/STS study confined to thin-film La$_{0.67}$Ca$_{0.33}$MnO$_{3}$ (LCMO), that is,
at optimal doping and therefore far from the regime where the material might become
intrinsically insulating. We vary thickness, strain, and surface morphology.
Strained films are grown on SrTiO$_{3}$ and unstrained films are grown on
NdGaO$_{3}$. Film thicknesses range from 10 to 180 nm. We expect the thinner films,
on the order of 10 to 25 nm, to be coherently strained and thicker films, of 100 nm
and more, to be strain-relaxed. We found that film surface morphology varied with
film thickness. Then we compare our surface-sensitive STM and STS measurements to
bulk-sensitive measurements of these same films.

%%%%%%%%%%%%%%%%%%%%%%%%%%%%%%%%%%%%%%%%%%%%%%%%%%%%%%%%%%%%%%%%%%%%%%%%%%%%%%%%%%%%%%
\section{Experiment}
%%%%%%%%%%%%%%%%%%%%%%%%%%%%%%%%%%%%%%%%%%%%%%%%%%%%%%%%%%%%%%%%%%%%%%%%%%%%%%%%%%%%%%

\subsection{Growth and Characterization}

Thin films of LCMO with thicknesses between 10 nm and 180 nm were grown on single
crystal substrates of SrTiO$_{3}$(001) and NdGaO$_{3}$(001). All films were grown
using dc sputtering from stoichiometric targets in 3 mbar of pure oxygen. Film
thickness was determined by sputtering time and verified with X-ray reflectivity
measurements of selected films (see below for details). Various substrate
temperatures were used during film growth, all between 780 $^\circ$C and 840
$^\circ$C. Some films were annealed \textit{in-situ}, immediately after sputtering.
Others were annealed \textit{ex-situ} in flowing pure oxygen at atmospheric
pressure.

Surface morphology was verified with tapping-mode atomic force microscopy.

Resistances versus temperature ($R$-$T$) transport measurements were made using four
in-line probes attached to the film with either silver paste or indium or using
structured bridges. For unstructured measurements, as the probes were of variable
size and with variable distance, only resistance and not resistivity was obtained.
Nevertheless, normalized $R$-$T$ curves are sufficient to observe the $M$-$I$
transition and CMR and to characterize film quality. X-ray diffraction (XRD)
measurements were made with a Siemens D5005 using a Cu K$\alpha_{1}$ source. Film
thicknesses were determined using low-angle XRD, while film quality was verified
with Laue oscillation measurements. X-ray photoemission spectroscopy (XPS)
measurements were made using a Mg anode (1253.6eV) and hemispherical analyzer in a
UHV measurement chamber with base pressure on the order of $10^{-10}$ mbar. Samples
were either baked overnight in a load-lock at approximately 100$^\circ$ C to desorb
surface contaminants, or brought into the measurement chamber rapidly to retain any
contamination present on the sample surface. Some films were plasma etched using
both O$_2$ and Ar at a calibrated etch rate of 1 nm/minute. This combination has
been shown to etch STO without depleting its oxygen and rendering it
metallic.\cite{Beekman2007}

\subsection{STM}

STM measurements were made under various conditions, ultra high vacuum (UHV),
flowing helium gas, and ambient. All of our STM heads were built in-house and use a
course approach based upon the Pan design.\cite{Wittneven1997} Both mechanically cut
PtIr (90\%:10\%) and electrochemically etched Pt wires were used as STM tips. The
UHV STM chamber has a base pressure of $4 \times 10^{-10}$ mbar, with the capability
to vary the sample temperature between 300 mK and 180K and the possibility to apply
magnetic fields up to 8 T. Samples were brought into the UHV chamber after being
pumped overnight in a load-lock.

Other STM measurements were conducted in helium boil-off gas within a variable
temperature insert (VTI) mounted through the bore of a 12 T magnet. Using resistance
heating, the sample temperature can be varied between 4.2 and 340 K. Once inserted
into the cryostat, samples were held between 300 K and 340 K while being flushed
with dry helium gas to effect desorption of contaminants. To minimize the sample
cryopumping while at low temperatures, samples were kept warmer than the surrounding
VTI. Ambient STM measurements were made using the same STM as used for helium gas
measurements, but without inserting the STM into the VTI.

Conductivity maps were measured using a digital lock-in amplifier. While scanning
topography, the STM bias voltage ($V_{bias}$) was modulated with an ac voltage
($V_{ac}$) from an oscillator built into the lock-in amplifier. The ac modulation was
limited in amplitude to a few percent of the bias voltage, and at a frequency higher
than the feedback bandwidth (on the order of 1 kHz). Conductivity was computed by the
lock-in using output from the STM current-to-voltage amplifier.

Current-voltage curves ($I$-$V$) were measured using a fixed tunneling gap method.
This begins with the STM in tunneling at particular bias ($V_{set}$) and current
($I_{set}$) set-points and with the feedback engaged. The STM feedback is then
disengaged, freezing the tip-sample distance while the bias voltage is swept through
the desired voltage range. Simultaneously, both the applied bias voltage and the
measured tunneling current were recorded. Up to 500 curves were taken and averaged for
each $I$-$V$ curve. Measured $I$-$V$ curves were numerically differentiated after
averaging. Resulting d$I$/d$V$ spectra were used as a proxy for sample density of
states (DOS).

%%%%%%%%%%%%%%%%%%%%%%%%%%%%%%%%%%%%%%%%%%%%%%%%%%%%%%%%%%%%%%%%%%%%%%%%%%%%%%%%%%%%%%
\section{Results}
%%%%%%%%%%%%%%%%%%%%%%%%%%%%%%%%%%%%%%%%%%%%%%%%%%%%%%%%%%%%%%%%%%%%%%%%%%%%%%%%%%%%%%

\subsection{Growth and Characterization}

\begin{figure} \centering \begin{tabular}{cc}

 \epsfig{file=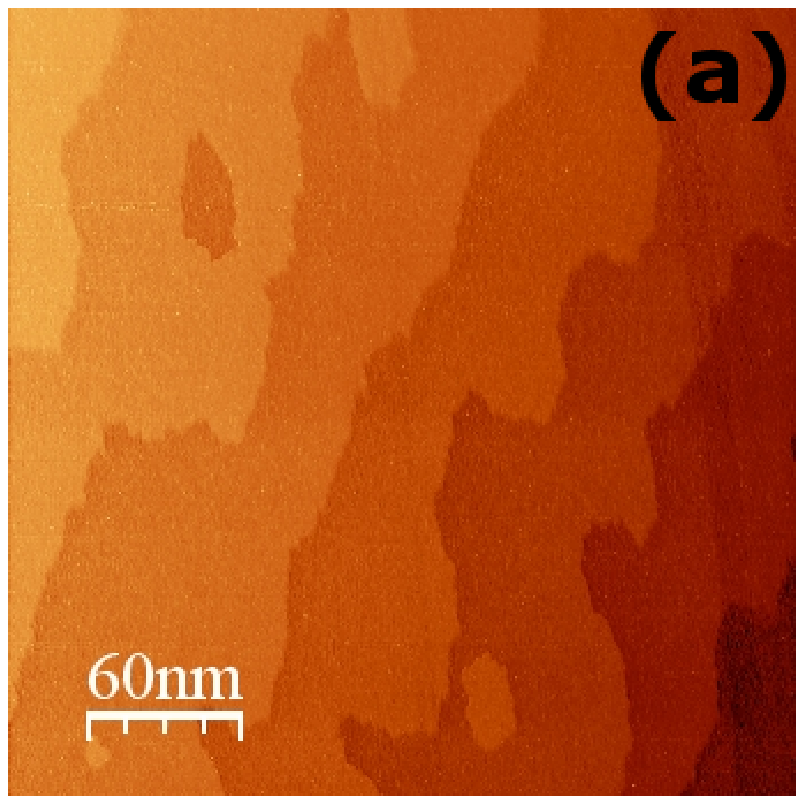,width=4.1cm,clip=} &

\epsfig{file=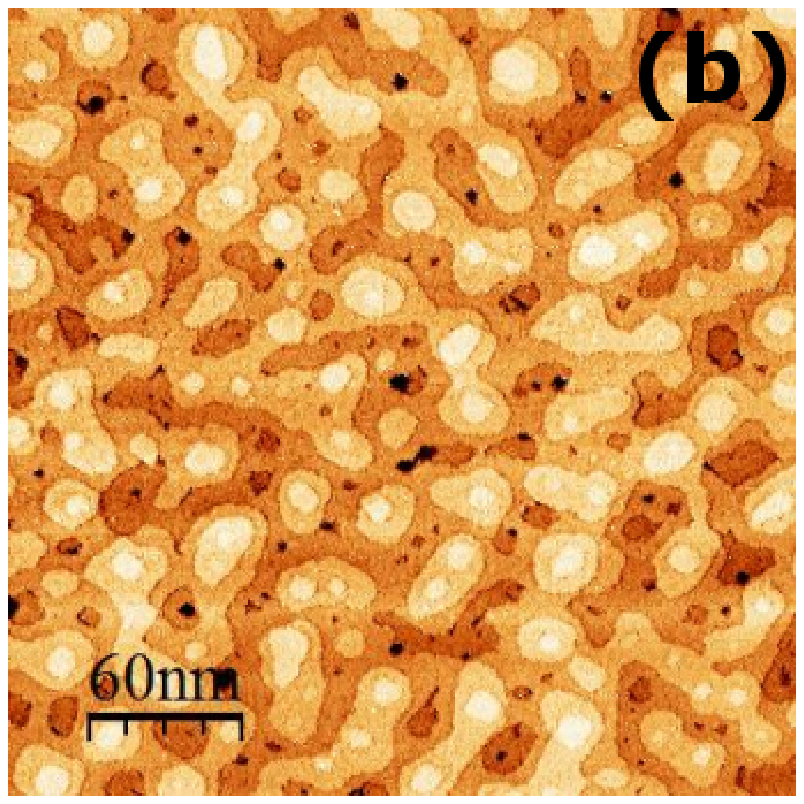,width=4.1cm,clip=} \end{tabular}

\caption{\label{STOmorph}(Color online) Ambient STM images of typical terraced, (a),
and island, (b), topograpy for La$_{0.67}$Ca$_{0.33}$MnO$_{3}$ grown on SrTiO$_{3}$.
Films are 10 nm thick. Both images exhibit unit-cell step heights.}

\end{figure}

With increasing film thickness, we observe that, for films grown on both NGO and STO,
the surface changes from flat terraces with a unit-cell step height, $\sim$0.4 nm, to a
rougher morphology, 10 to 20 nm peak-to-peak, with no indication of terraces. We also
observe that terraced films can be further divided into those with long narrow terraces
and those exhibiting 2-D island growth. Typical examples of terraced and 2-D island
growth morphology are shown in Figure \ref{STOmorph}. Both of these films are 10 nm
thick, were grown on STO substrates, and measured with an STM in ambient conditions.
Although for films grown on STO with thicknesses between 10 and 50 nm both morphologies
were found, the predominant morphology was terraced. Since growth conditions were
nominally the same, we attribute the occasional appearance of island growth to the
varying quality of the commercial substrates. However, other researchers have noted a
change from terrace growth to 2-D island growth for LCMO grown on low miscut angle STO
as films thickness is increased above 30 nm.\cite{Sanchez2006} For our thicker films on
STO, those about 100 nm and thicker, the surface morphology becomes rough. Terraced and
2-D island growth was also seen for films grown on NGO and are similar in appearance to
the STO films shown in Figure \ref{STOmorph}. However, for films grown on NGO, a
correlation between film thickness and surface morphology is apparent. The thinnest
films, around 10 nm thick, were almost exclusively terraced. Medium thickness films,
around 25 nm thick, generally exhibited 2-D island growth. Thicker films, those thicker
than 50 nm, were usually rough.

\begin{figure}

\epsfig{file=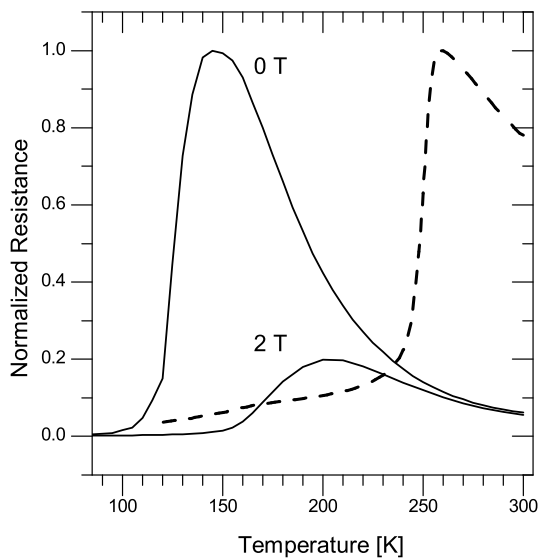,width=9cm,clip=}

\caption{\label{RT}Resistance versus temperature curves normalized to the zero-field
peak resistance. The solid curve is of a 10 nm La$_{0.67}$Ca$_{0.33}$MnO$_{3}$ film
grown on SrTiO$_{3}$, measured in the indicated magnetic field ($T_p\approx145$K).
The dashed curve is of a 26 nm La$_{0.67}$Ca$_{0.33}$MnO$_{3}$ film grown on
NdGaO$_{3}$ ($T_p\approx260$ K).}

\end{figure}

Transport measurements were made on some of our films. Shown in Figure \ref{RT} are
example $R$-$T$ measurements for flat LCMO films grown on STO and NGO. From such
curves we can extract the peak temperature ($T_p$). These curves were normalized to
the zero-field peak resistance. The peak temperatures are typical for strained films
grown on STO ($T_p\approx145$ K) and unstrained films grown on NGO ($T_p\approx260$
K), though $T_p$ does vary a little with film thickness. Also shown for the STO film
is an $R$-$T$ measurement in magnetic field that confirms this film to be
magnetoresistive. Similar measurements were made of other films, both flat and
rough. We can conclude that the film bulk of both our flat and rough films have an
$M$-$I$ transition and are magnetoresistive.

%+++++++++++++++++++++++++++++++++++++++++++++++++++++++++++++++++++++++++++++++++++++
%+++++++++++++++++++++++++++++++++++++++++++++++++++++++++++++++++++++++++++++++++++++

%+++++++++++++++++++++++++++++++++++++++++++++++++++++++++++++++++++++++++++++++++++++
%+++++++++++++++++++++++++++++++++++++++++++++++++++++++++++++++++++++++++++++++++++++

XPS measurements were made of a subset of films to confirm the presence of lanthanum,
calcium, manganese, and oxygen and to verify the absence of other, contaminating
elements at the film surface. Except for carbon, no evidence of contamination was
observed. Since carbon contamination is ubiquitous, and as sample cleaning was limited
to moderate \textit{in-situ} heating, the presence of carbon is not unexpected, and
does not necessarily indicate a poor quality film.\cite{Beyreuther2006}

%%%%%%%%%%%%%%%%%%%%%%%%%%%%%%%%%%%%%%%%%%%%%%%%%%%%%%%%%%%%%%%%%%%%%%%%%%%%%%%%%%%%%%
\subsection{STM}
%%%%%%%%%%%%%%%%%%%%%%%%%%%%%%%%%%%%%%%%%%%%%%%%%%%%%%%%%%%%%%%%%%%%%%%%%%%%%%%%%%%%%%

STM and STS measurements were taken of LCMO films with various thicknesses grown on STO
and NGO substrates. Both flat films, those with terraced or 2-D island growth, and
rough films, those with rounded morphology and no discernable unit-cell steps, were
measured.

\subsubsection{Rough Morphology}

%+++++++++++++++++++++++++++++++++++++++++++++++++++++++++++++++++++++++++++++++++++++
\begin{figure} \centering \begin{tabular}{cc}

 \epsfig{file=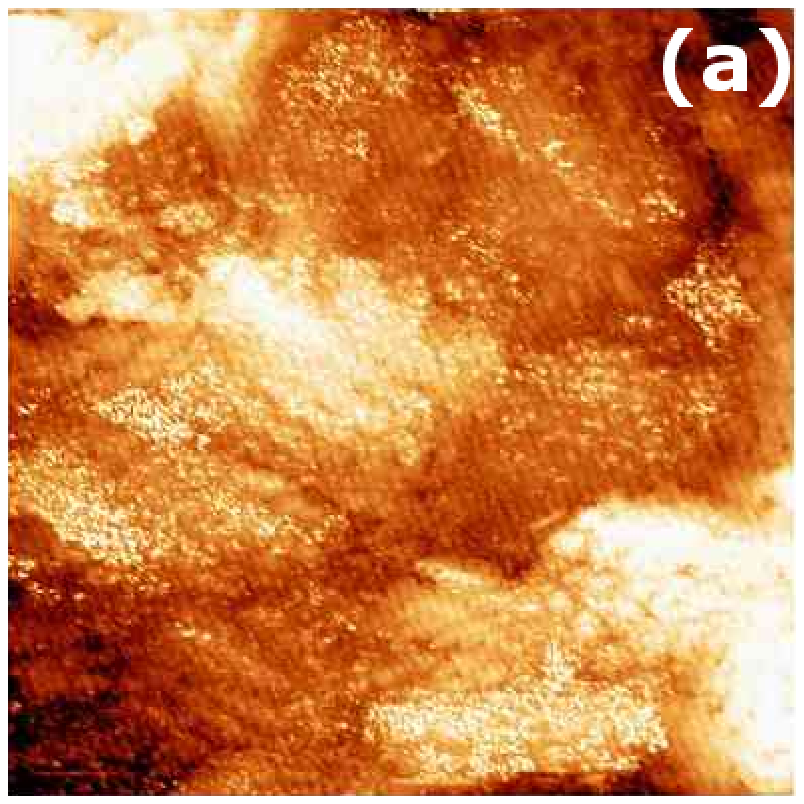,width=4.1cm,clip=} &

 \epsfig{file=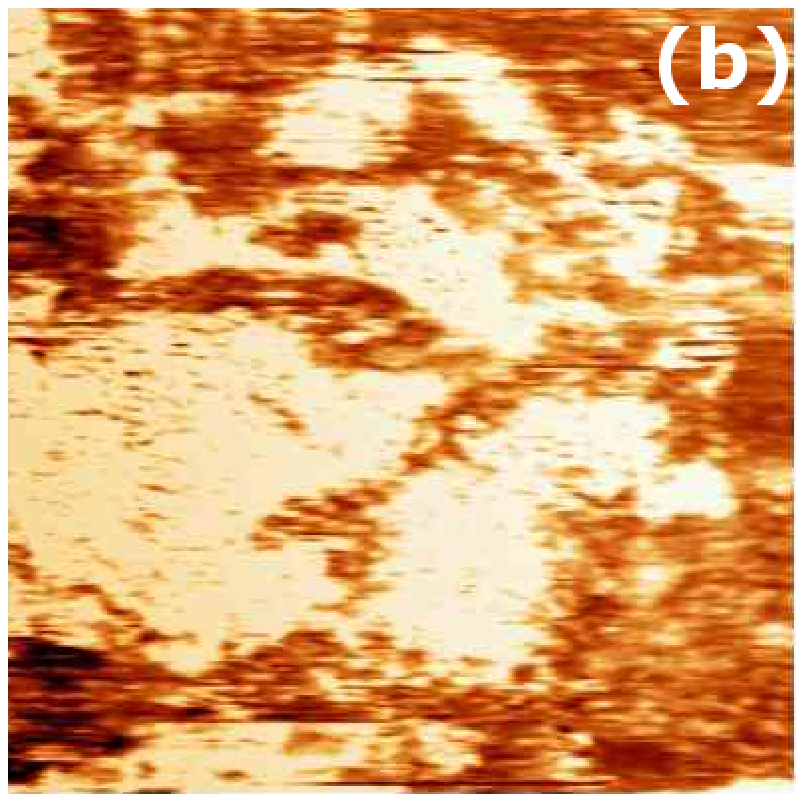,width=4.1cm,clip=} \\

 \epsfig{file=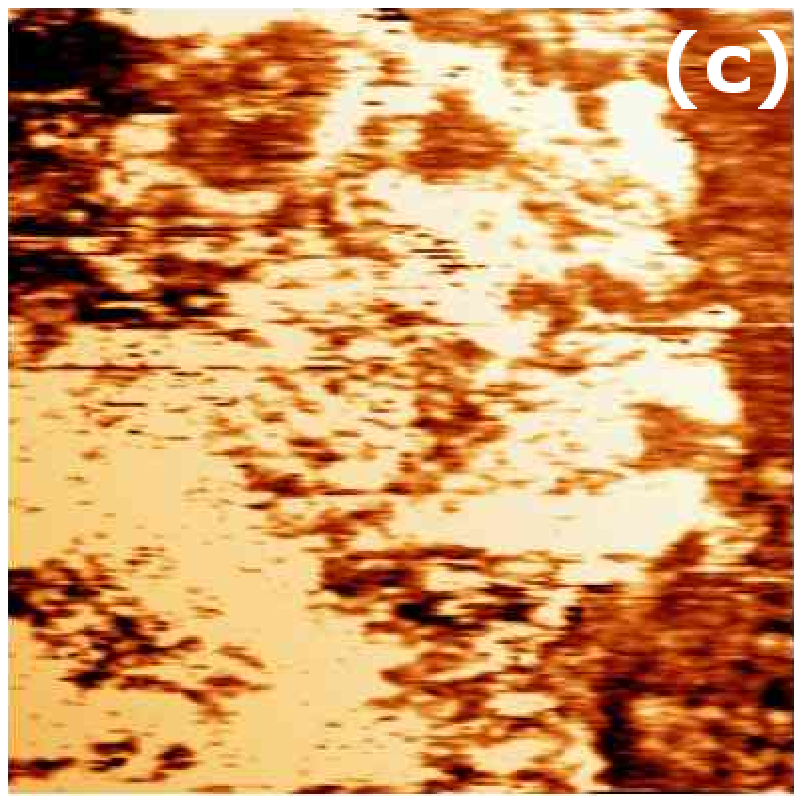,width=4.1cm,clip=} &

 \epsfig{file=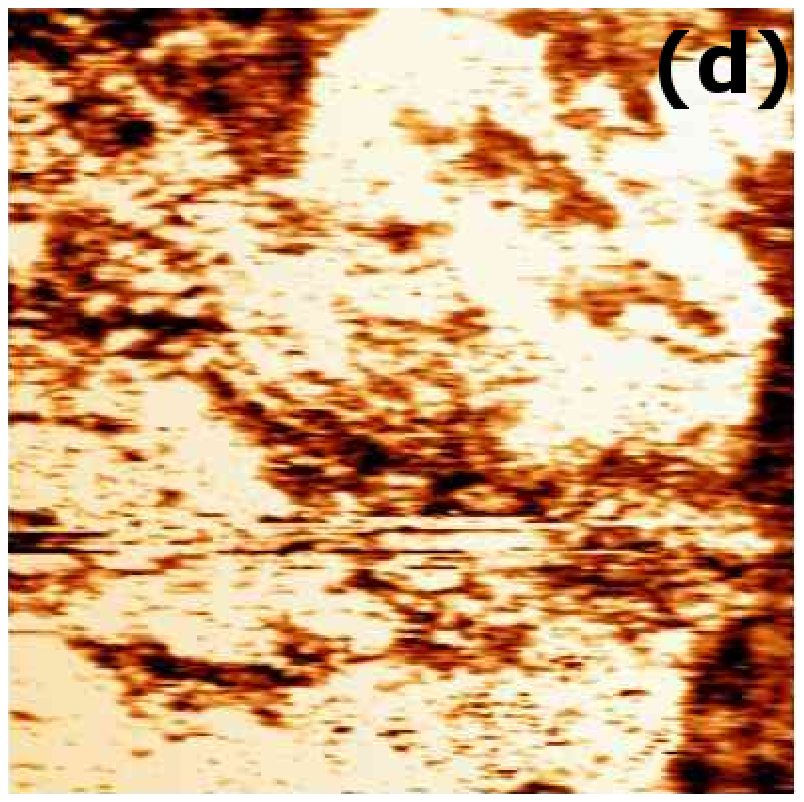,width=4.1cm,clip=} \\

 \epsfig{file=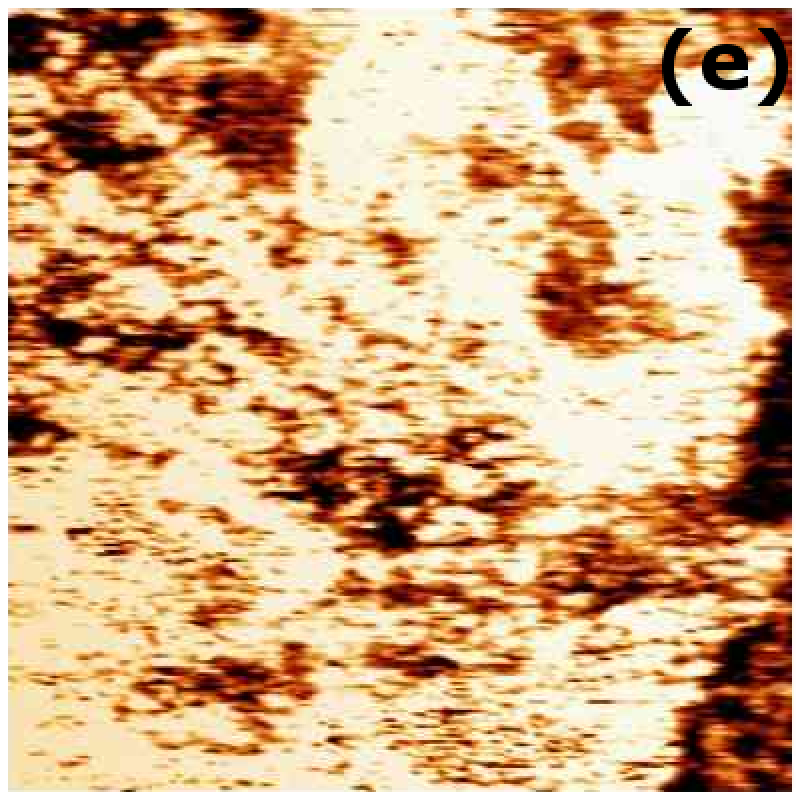,width=4.1cm,clip=} &

 \epsfig{file=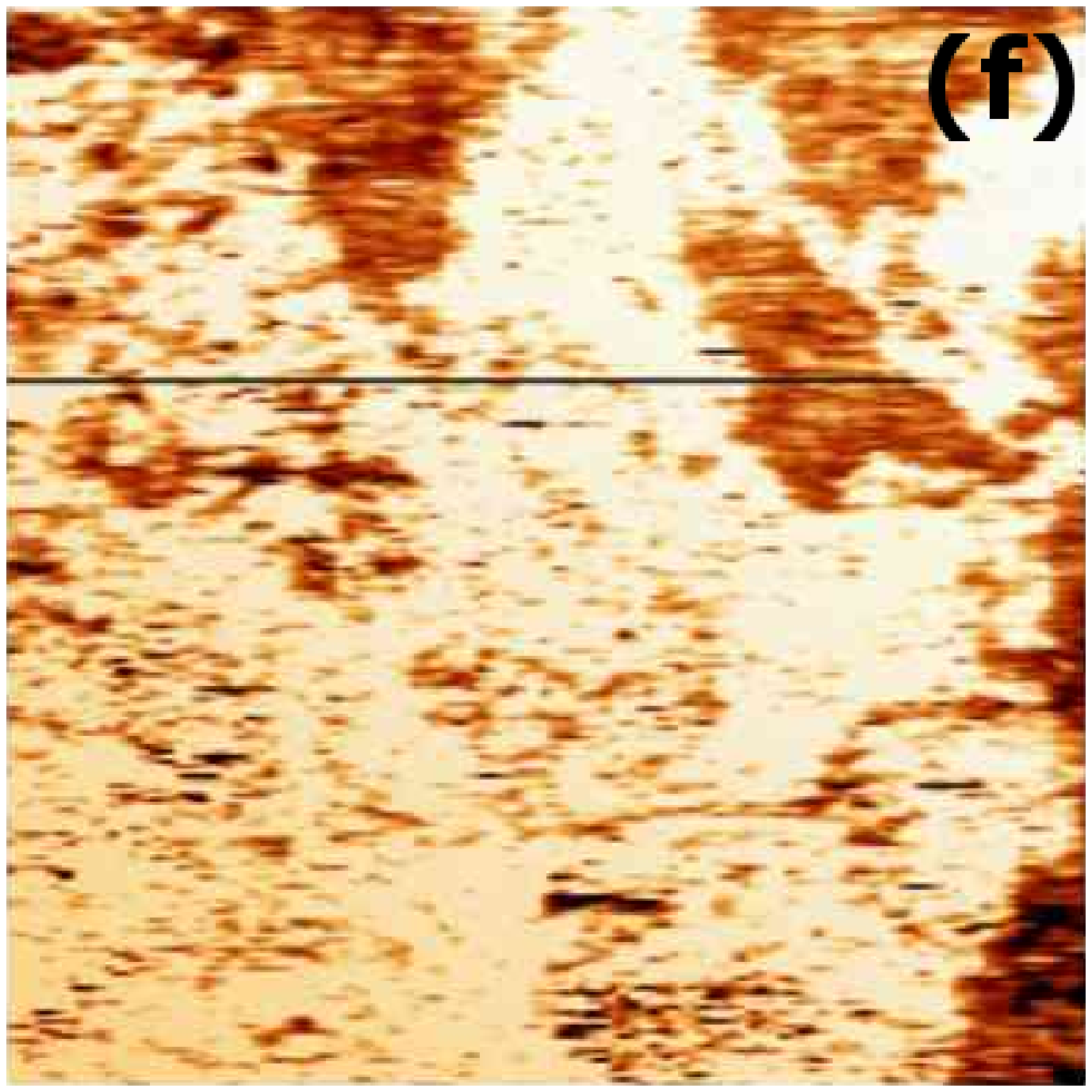,width=4.1cm,clip=}

 \end{tabular}

\caption{\label{CMR}(Color online) STM image of rough 100 nm thick
La$_{0.67}$Ca$_{0.33}$MnO$_{3}$ film grown on Nb-doped SrTiO$_{3}$ substrate. All
images are 500 nm$^{2}$. Topography (20 nm peak-to-peak) is shown in frame (a).
Conductivity maps with applied magnetic fields of 0, 2, 4, 6, and 7 Tesla are shown
in (b), (c), (d), (e), and (f) respectively. Darker areas are more metallic and
lighter areas are more insulating. Measurements were in helium gas at 50K,
$V_{set}$=-2.0 V, $I_{set}$=0.2 nA, and $V_{ac}$=24 mV.}

\end{figure}
%+++++++++++++++++++++++++++++++++++++++++++++++++++++++++++++++++++++++++++++++++++++

Thicker LCMO films grown on either STO or NGO were rough and show evidence of phase
separation and an electronically and magnetically active surface layer. Figure
\ref{CMR} (a) displays topography for a 100 nm thick film grown on Nb-doped STO. The
film surface is rough, with peak-to-peak height variations of 20 nm. Also seen in
Figure \ref{CMR} is a sequence of conductivity maps covering the same area as shown in
Figure \ref{CMR} (a), but for different applied magnetic fields. As the applied field
is increased, the metallic (dark) area increases and the insulating (light) area
decreases. Note that some insulating regions do increase in size at higher fields.
Nevertheless, the predominant trend is an overall increase in metallic regions. There
is some correlation between surface morphology and conductivity, though not as
significantly as seen in other films (see below). Although these conductivity maps were
measured at 50 K, well below the $M$-$I$ transition temperature, electronic
inhomogeneities are still present, even with a 7 Tesla applied magnetic field. Similar
results were obtained by F\"ath \textit{et al},\cite{Fath1999} on a similarly rough
LCMO film, although their film was grown with a YBa$_{2}$Cu$_{3}$O$_{7}$ layer between
the LCMO film and the STO substrate.

%+++++++++++++++++++++++++++++++++++++++++++++++++++++++++++++++++++++++++++++++++++++
\begin{figure} \centering \begin{tabular}{cc}
\epsfig{file=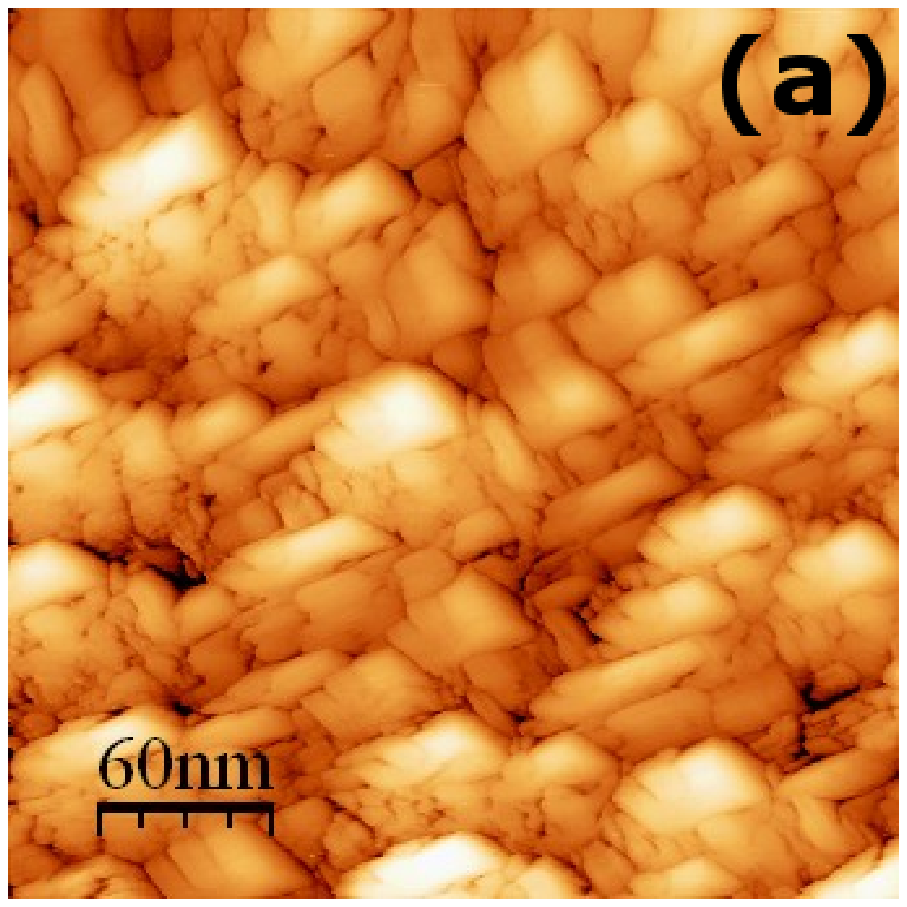,width=4.1cm,clip=} &
\epsfig{file=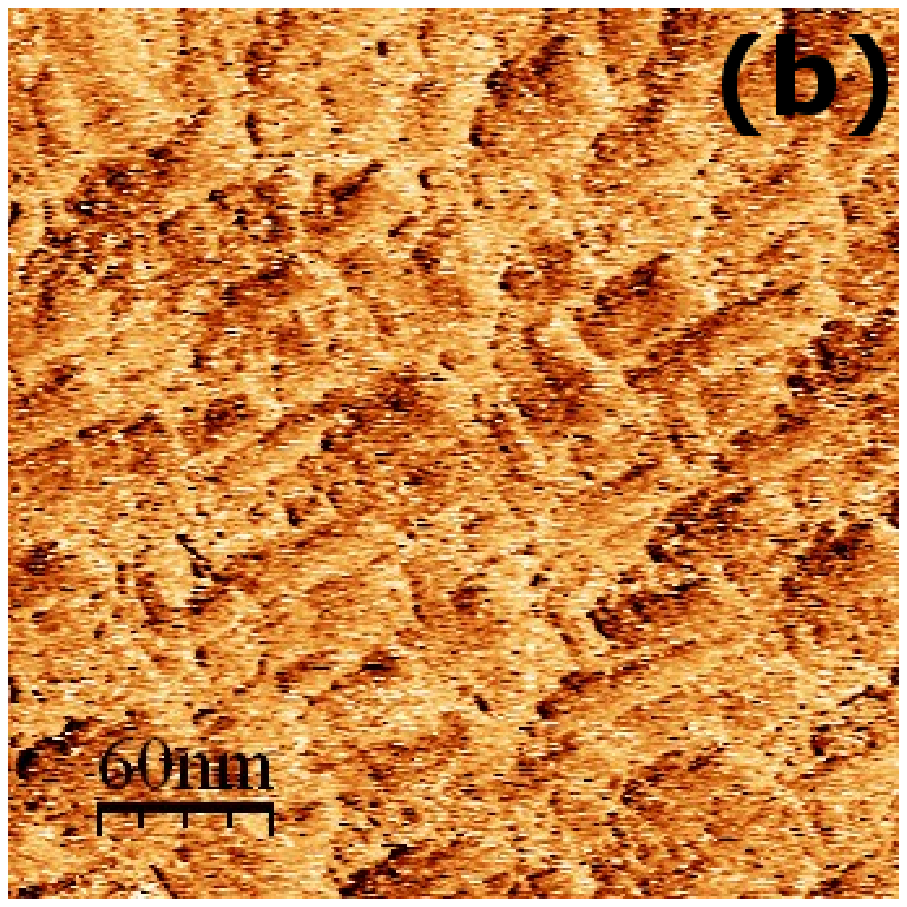,width=4.1cm,clip=}
 \end{tabular}

\caption{\label{NGOrough}(Color online) STM image of a rough 50 nm thick
La$_{0.67}$Ca$_{0.33}$MnO$_{3}$ film grown on NdGaO$_{3}$. (a) Topography and (b)
conductivity map. Measurements were in helium gas at 280 K with zero magnetic field.
$V_{set}$=-0.5 V, $I_{set}$=0.5 nA, and $V_{ac}$=10 mV.}

\end{figure}
%+++++++++++++++++++++++++++++++++++++++++++++++++++++++++++++++++++++++++++++++++++++

%+++++++++++++++++++++++++++++++++++++++++++++++++++++++++++++++++++++++++++++++++++++
\begin{figure} \centering

\epsfig{file=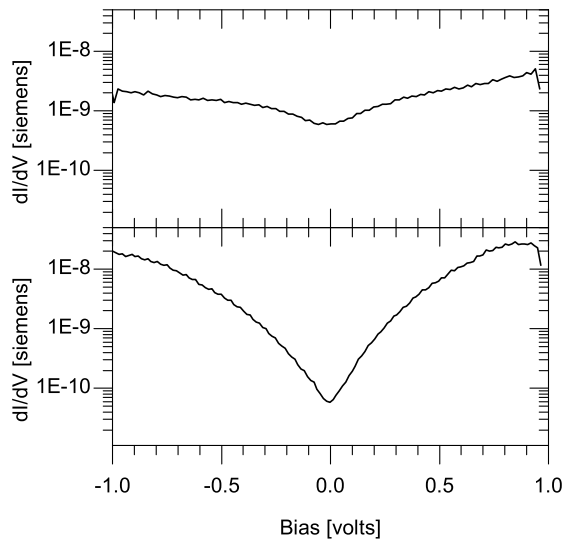,width=9cm,clip=}

\caption{\label{NGOroughdIdV}Numerically differentiated $I$-$V$ curves of a rough 50
nm thick La$_{0.67}$Ca$_{0.33}$MnO$_{3}$ film grown on NdGaO$_{3}$, measured at 299
K on different locations on the film. V$_{set}$=-1.0V, I$_{set}$=0.5nA, $\pm$1V.
Same film as shown in Figure \ref{NGOrough}}

\end{figure}
%+++++++++++++++++++++++++++++++++++++++++++++++++++++++++++++++++++++++++++++++++++++

Rough films grown on NGO were also inhomogeneous. This is confirmed by both
conductivity maps and d$I$/d$V$ spectra. An example of topography and conductivity map
of a rough LCMO film on NGO is shown in Figure \ref{NGOrough}. This film is 50 nm
thick, with rough topography (10 nm peak-to-peak), and, as with rough samples on STO,
with an inhomogeneous conductivity map (darker regions are more metallic, lighter
regions are more insulating). Note that there is significant correlation between the
conductivity map and topography. This is not because of feedback-error since the
forward and reverse scans of the conductivity map are very similar, but rather reflect
properties of the surface. This particular measurement was made at 280 K, just above
$T_p$, where phase separation would be expected. However, other conductivity maps of
this film recorded above and below $T_p$ are similarly inhomogeneous. Unfortunately,
due to the difficulty of calibrating conductivity maps measured with a lock-in
amplifier, only qualitative information can be gleaned from such measurements. To
provide quantitative information, d$I$/d$V$ spectra were also taken of this film.
Spectra taken at different locations indicate the presence of both metallic and
insulating regions, confirming the electronic inhomogeneity seen in conductivity maps.
Examples of d$I$/d$V$ spectra are shown in Figure \ref{NGOroughdIdV}. Both curves were
measured on the same film as shown in Figure \ref{NGOrough} but at 299 K, significantly
above $T_p$. The zero bias conductivity at this temperature varies by an order of
magnitude, from $6.5 \times 10^{-11}$ to $6.1 \times 10 ^{-10}$ siemens. These curves
demonstrate that even well above $T_p$, rough films are inhomogeneous. Other d$I$/d$V$
spectra measured in and well below the $M$-$I$ transition indicate electronic
inhomogeneity. Measurements of other rough films confirm these results.

\subsubsection{Smooth Morphology}

\begin{figure} \centering \begin{tabular}{cc}
\epsfig{file=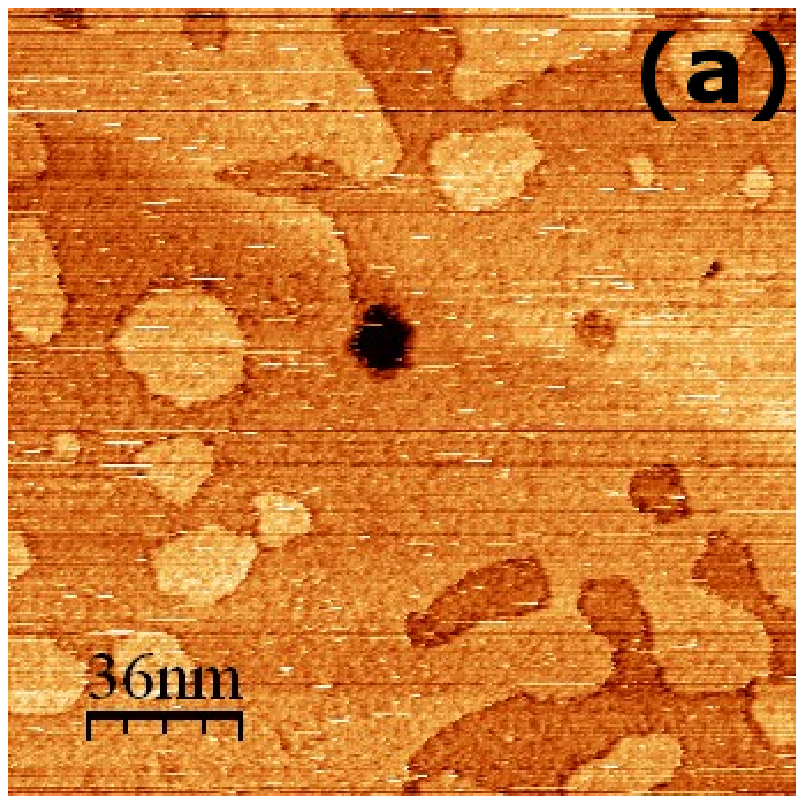,width=4.1cm,clip=} &
\epsfig{file=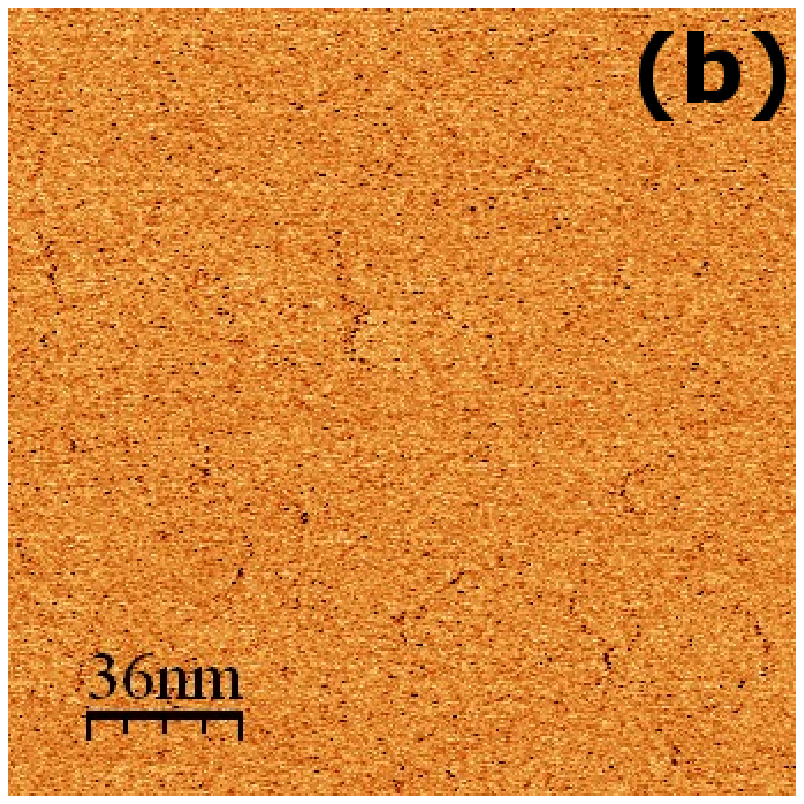,width=4.1cm,clip=}
 \end{tabular}

\caption{\label{STOcond}(Color online) STM measurements of atomically smooth 10 nm
thick La$_{0.67}$Ca$_{0.33}$MnO$_{3}$ grown on SrTiO$_{3}$. (a) Topography and (b)
conductivity map, measured in helium gas at 150K with zero magnetic field. The
height between terraces matches unit-cell size.}

\end{figure}

Thinner films grown on either NGO or STO were atomically smooth and electronically
homogeneous with electronically and magnetically inactive surface layers. A typical
example of topography and conductivity for a 10 nm thick terraced STO film is shown in
Figure \ref{STOcond}. This measurement was made in zero magnetic field at 150 K, close
to the film's $T_p$. For a first-order phase transition, we expect to see
inhomogeneities at temperatures close to $T_p$. However, we saw no evidence of phase
separation at this temperature or at any other. What detail there is in this
conductivity map is at terrace edges and defects. We attribute this to feedback error,
the increased or decreased conductivity resulting from the tip coming closer to or
going further away from the surface. Varying the STM bias voltage, current set point,
or ac-modulation voltage did not change this result. Conductivity maps taken at
temperatures above and below the $M$-$I$ transition were similarly featureless.
Conductivity maps were also made of other flat LCMO films on STO. All were similarly
homogeneous, regardless of the temperature. Similar results were seen for smooth films
grown on NGO.

%+++++++++++++++++++++++++++++++++++++++++++++++++++++++++++++++++++++++++++++++++++++
%+++++++++++++++++++++++++++++++++++++++++++++++++++++++++++++++++++++++++++++++++++++

%+++++++++++++++++++++++++++++++++++++++++++++++++++++++++++++++++++++++++++++++++++++
\begin{figure}

\epsfig{file=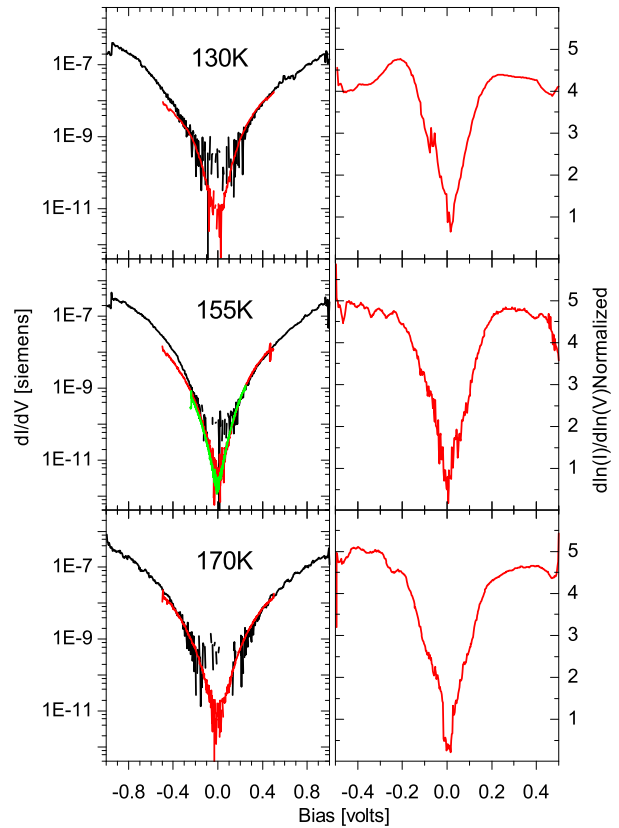,width=9cm,clip=}

\caption{\label{STOdIdV}(Color) Numerically differentiated $I$-$V$ curves of a
smooth 52 nm thick La$_{0.67}$Ca$_{0.33}$MnO$_{3}$ film grown on SrTiO$_{3}$.
Left-hand column d$I$/d$V$ with $V_{set}$=1.0 V, $\pm$1.0 V (black); $V_{set}$=0.5
V, $\pm$0.5 V (red); and $V_{set}$=0.25 V, $\pm$0.25 V (green). Right-hand column
normalized d$I$/d$V$ $V_{set}$=0.5 V, $\pm$0.5 V. All at $I_{set}$=0.5 nA. Measured
in UHV at indicated temperature with zero magnetic field.}

\end{figure}
%+++++++++++++++++++++++++++++++++++++++++++++++++++++++++++++++++++++++++++++++++++++

$dI$/$dV$ spectra were taken of flat films grown on STO and NGO. These spectra show
small variations with temperature, but no consistent trend in zero-bias
conductivity. They also indicate that our flat films were spatially homogeneous. An
example of d$I$/d$V$ spectra measured on a terraced STO film is shown in the
left-hand column of Figure \ref{STOdIdV}. For this film, 52 nm thick LCMO on STO,
$T_p \approx$155 K. Shown in the figure are spectra from above, below, and at $T_p$.
At each temperature are plotted the spectra calculated from two (or three) $I$-$V$
measurements at bias set-points 0.25 (green), 0.5 (red), and 1.0 V (black), but all
with $I_{set}$=0.5 nA. In these $I$-$V$ measurements, the bias was swept from the
$V_{set}$ to $-V_{set}$. The spectra measured at $V_{set}=0.25 V$ and $V_{set}=0.5
V$ were linearly scaled by a procedure explained in the Appendix. The need to
measure at different set-points is made clear by looking at the data for
$V_{set}=1.0 V$. Between $\pm$0.3~V, the spectra are very noisy because between
these voltages the resulting current is below the system noise floor. Reducing
$V_{set}$ while maintaining $I_{set}$ fixed increases the set-point conductance
linearly, but, because of the non-linear nature of the tunneling gap, the
conductance at low voltages has increased much more. This point is expanded upon in
the Appendix. Because of this increase in conductance, and as can be seen at all
three temperatures, more of the narrow spectra are above the noise floor. From this
it must be concluded that there is not a gap, but only a depletion of the DOS at
zero bias. At other temperatures, varying between 170 K and 80 K, d$I$/d$V$ spectra
of this film also show no significant variation. Other smooth LCMO films grown on
STO and NGO were measured and also demonstrate no consistent variation in d$I$/d$V$
spectra with temperature.

Shown in the right-hand column of Figure \ref{STOdIdV} are the d$I$/d$V$ spectra
from the left-hand column at $V_{set}$=0.5~V , but normalized by dividing by $I/V$.
Normalizing in this way attempts to extract the DOS from the d$I$/d$V$ spectra, by
canceling the effect of the tunneling gap.\cite{Stroscio1993} With this
normalization procedure, earlier studies claimed to observe both polaron
peaks\cite{Wei1997,Seiro2008} and half-metallicity\cite{Wei1997} in manganite films.
We measured d$I$/d$V$ spectra from many flat films grown on both STO and NGO and
normalized these spectra with this method. Some of our data did suggest polaron
peaks, but these polaron peaks do not appear on all films measured, even if grown
under identical process conditions. When polaron peaks were evident, the
peak-to-peak gap varied little with temperature, and no trend, such as seen in
Ref.~\cite{Seiro2008} was apparent. We applied this normalization procedure to
d$I$/d$V$ spectra taken at larger voltage setpoints ($V_{set} \geq 1.0$ volt), but
found that, in the range of bias voltages where we would expect to see polaron
peaks, the signal-to-noise ratio was too low.

%+++++++++++++++++++++++++++++++++++++++++++++++++++++++++++++++++++++++++++++++++++++
\begin{figure} \centering

\epsfig{file=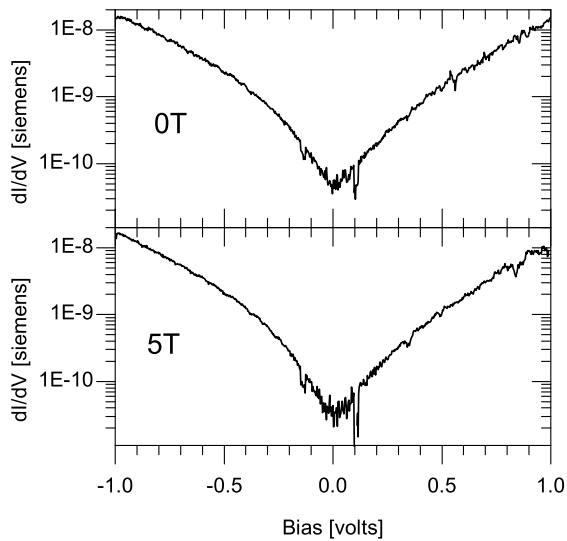,width=9cm,clip=}

\caption{\label{fielddIdV}Numerically differentiated $I$-$V$ curves of a smooth 26
nm thick La$_{0.67}$Ca$_{0.33}$MnO$_{3}$ film grown on NdGaO$_{3}$. Measured in
helium gas at 265K with 0 and 5 Tesla applied magnetic field. V$_{set}$=-0.7V,
I$_{set}$=1.0nA, $\pm$1V.}

\end{figure}
%+++++++++++++++++++++++++++++++++++++++++++++++++++++++++++++++++++++++++++++++++++++

We also measured atomically smooth films in an applied magnetic field. Shown in
Figure \ref{fielddIdV} are d$I$/d$V$ spectra measured on a smooth 26~nm thick film
grown on NGO in zero field and in a 5 Tesla field. This data was taken at T=265~K,
close to the $M$-$I$ transition of this film ($T_p$=260~K). These spectra are very
similar to each other and to other spectra taken on other parts of the film, and
also to spectra taken at other temperatures. Any differences are attributable to
experimental error and show no correlation with either temperature or magnetic
field. As we expect a large negative magnetoresistance in the vicinity of the
$M$-$I$ transition, this result suggests that the film surface is magnetically
inactive. Transport measurements (shown in Figure \ref{RT}) verify that the bulk of
this film does go through a $M$-$I$ transition. Again, this suggests the film
surface is inactive. Other smooth films on NGO and STO were similarly unresponsive
to magnetic fields.

%%%%%%%%%%%%%%%%%%%%%%%%%%%%%%%%%%%%%%%%%%%%%%%%%%%%%%%%%%%%%%%%%%%%%%%%%%%%%%%%%%%%%%
\subsubsection{Annealing and Etching}
%%%%%%%%%%%%%%%%%%%%%%%%%%%%%%%%%%%%%%%%%%%%%%%%%%%%%%%%%%%%%%%%%%%%%%%%%%%%%%%%%%%%%%

We attempted to remove the electronically inactive layer observed on flat films
using a plasma etcher. Similarly to Hudson \textit{et al},\cite{Hudson2001} we found
that once etched, the film surface became electronically active, but that the
results were inconsistent from sample to sample. Etching 1 or 2 nm did modify the
d$I$/d$V$ spectra, but did not result in the appearance of metallic spectra (meaning
a significant increase in zero-bias conductivity). After etching about 3 nm, we were
able to measured some metallic d$I$/d$V$ spectra. On these films we typically
measured a number of metallic d$I$/d$V$ spectra at temperatures somewhat below
$T_p$. The typical temperature window for this metallic phase was between 5 and 15 K
wide. However, upon cooling to lower temperatures, the metallic phase was no longer
evident, and instead of staying metallic, the d$I$/d$V$ spectra became insulating
and similar to those seen above the transition. Atomic force microscopy images of
etched films show that etching initially followed the existing topography, leaving
terraces with an appearance similar to as-grown. Further etching (3 nm or more)
rounded off terrace edges, roughening the film, and giving it the appearance of an
as-grown rough film. Etching did not produce spectra consistent with the $M$-$I$
transition seen in transport measurements of these films, and it is unlikely that
our STS data are representative of film-bulk behavior. The results also indicate
that etching does something else than simply removing an inactive surface layer from
the active bulk material.

\begin{figure} \centering \begin{tabular}{cc}

\epsfig{file=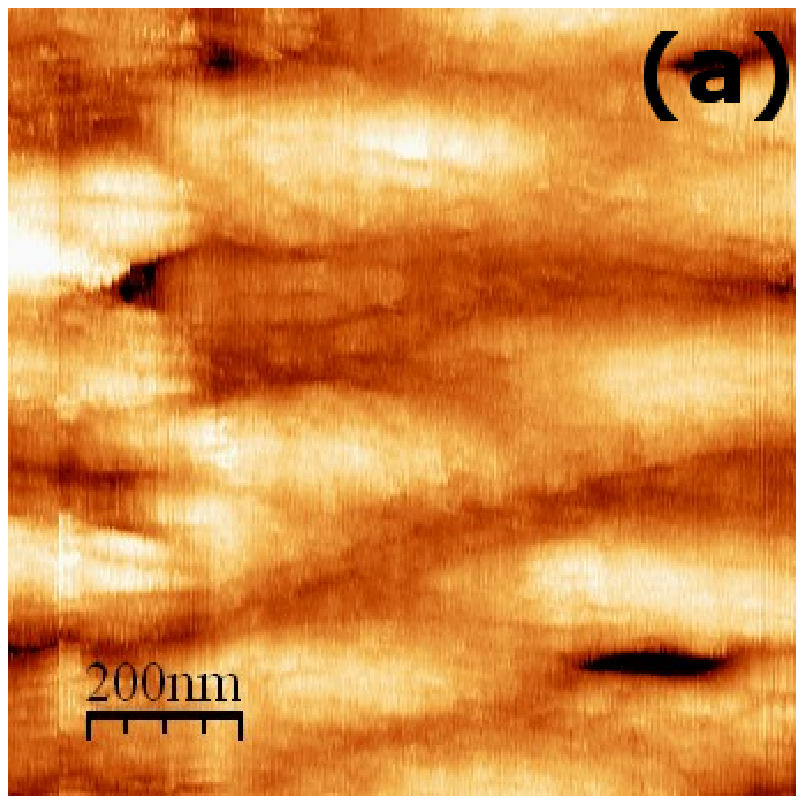,width=4.1cm,clip=} &
 \epsfig{file=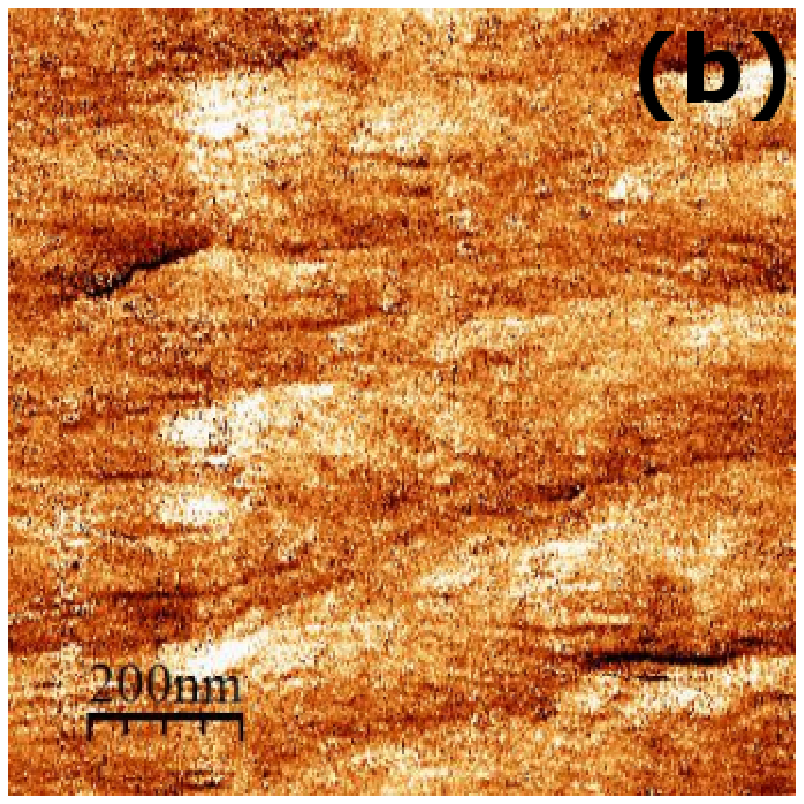,width=4.1cm,clip=}
 \end{tabular}

\caption{\label{NGOannealed}(Color online) STM image of an annealed 26 nm thick
La$_{0.67}$Ca$_{0.33}$MnO$_{3}$ film grown on NdGaO$_{3}$ after annealing at
470$^\circ$C for 3 hours in flowing oxygen. (a) Topography and (b) conductivity map.
Measured in ambient.}

\end{figure}

One possible explanation for the inactive surface layer is oxygen depletion. We
attempted to reoxidize films by annealing them in an oxygen-rich atmosphere. One
sample was annealed \textit{in-situ} in 1000 mbar of oxygen at 650$^\circ$C for 30
minutes. This sample proved to be electronically inactive. We also tried to eliminate
possible causes of oxygen depletion. Although our films were typically grown in 3 mbar
of oxygen, our usual process recipe called for evacuating the process chamber
immediately after sputtering was complete. Thus while cooling, the film would be
exposed a vacuum on the order of $1 \times 10^{-6}$ mbar. Beyreuther \textit{et al}
found below-stoichiometry oxygen in manganite films exposed to UHV at elevated
temperatures (470-670$^\circ$C),\cite{Beyreuther2006} and so we attempted to minimize
film deoxidization by maintaining 3 mbar of oxygen during cooling. The resulting film
was still electronically inactive. Following their recipe for film reoxidization, we
annealed a sample at 470$^\circ$C in flowing oxygen for 3 hours. The resulting film
was electronically active, but inhomogeneous. Ambient STM topography and conductivity
map shown in Figure \ref{NGOannealed}. The film was roughened by the annealing. In
contrast, before annealing, the film exhibited 2-D island growth with a homogeneous
conductivity map.

%+++++++++++++++++++++++++++++++++++++++++++++++++++++++++++++++++++++++++++++++++++++
%+++++++++++++++++++++++++++++++++++++++++++++++++++++++++++++++++++++++++++++++++++++

%%%%%%%%%%%%%%%%%%%%%%%%%%%%%%%%%%%%%%%%%%%%%%%%%%%%%%%%%%%%%%%%%%%%%%%%%%%%%%%%%%%%%%
\section{Discussion}
%%%%%%%%%%%%%%%%%%%%%%%%%%%%%%%%%%%%%%%%%%%%%%%%%%%%%%%%%%%%%%%%%%%%%%%%%%%%%%%%%%%%%%

We were able measure electronically active surface layers on rough films. Other
researchers have found similar results using STM/STS on rough films. The surface of
our flat, smooth films are electronically and magnetically inactive. However, using
methods other than STM/STS, researchers have also observed inactive surface layers
on manganites, and a number of explanations for the origin of this layer have been
put forth. Measurements have detected both oxygen depletion and dopant enrichment at
the surface, both of which would change its electronic structure. Others have
proposed surface electronic reconstruction driven by symmetry breaking at the
crystal surface. We now discuss these various points in some detail.

\subsection{Other STM Results}
\subsubsection{Rough Films}

In agreement with our data, other researchers have measured inhomogeneities on rough
films. Becker \textit{et al} measured an LCMO film grown on MgO and found it to be
electronically inhomogeneous.\cite{Becker2002} Their film was 200 nm thick, with a
rough, grain-like surface morphology. F\"ath \textit{et al} detected inhomogeneities
on a rough (maximum 20nm peak-to-peak roughness) LCMO film on STO, and measured an
increase in metallicity with increased magnetic field, as would be expected of a CMR
material.\cite{Fath1999} Chen \textit{et al} also detected electronic
inhomogeneities on an LCMO film grown on STO.\cite{Chen2003} Their film was 300 nm
thick with peak-to-peak height variation of 7 nm over 50 nm lateral distance, with
no evidence of terraces or unit-cell steps.

\subsubsection{Atomically Flat Films}

Although a number of researchers {\it reported} electronically active surface layers
on flat LCMO films using STM/STS, taken collectively, no clear picture emerges.
Seiro \textit{et al} published STS measurements of an LCMO film on
STO.\cite{Seiro2008} STM topography indicated a terraced surface, and STS
measurements showed the film to be spatially electronically homogeneous. They
observe conductance peaks in their normalized d$I$/d$V$ spectra, and claim these
peaks to be the signature of polarons, with half the peak-to-peak distance
representing the polaron binding energy. They found this binding energy varied with
temperature, but narrowed in the region of $T_{p}$, exactly where their transport
measurements show the film to have its \textit{highest} resistivity. Other groups
also measured flat, but electronically active films. Both Biswas \textit{et
al}\cite{Biswas1999} and Mitra \textit{et al}\cite{Mitra2005} measured atomically
flat LCMO films and observe that, in the vicinity of $T_{p}$, a gap in the d$I$/d$V$
spectra opens, corresponding to the peak resistivity of transport measurements.
Conductivity maps measured by Mitra \textit{et al} show their film to be
electronically homogeneous, and they attribute this to their film being grown on NGO
and therefore being unstrained. However, Biswas \textit{et al} measured strained
LCMO grown on LAO and make no mention of electronic inhomogenities. Both Seiro
\textit{et al}\cite{Seiro2008} and Mitra \textit{et
al}\cite{Mitra2005,Paranjape2003} appear to present data taken from only one sample.
We found inconsistencies between our films grown under nominally identical process
conditions, and found our conclusions to be more robust when drawn from a number of
samples. Furthermore, both groups present example $I$-$V$ curves that appear to be
measured at different voltage set points. Considering the difficulty of separating
sample DOS from the tunneling barrier,\cite{Ukraintsev1995, Koslowski2007} we would
not want to draw conclusions from such a procedure and prefer to compare data taken
at a single set-point, eliminating one source of uncertainty.

Other than our results presented here, only one other group has presented STM/STS data
showing conductivity maps of both flat and rough films. Moshnyaga \textit{et al}
measured LCMO films on MgO at 115 K, significantly below the film $M$-$I$ transition
temperature.\cite{Moshnyaga2006} They found their flat film to be electronically
homogeneous, while their rough film was inhomogeneous. Magnetization measurements
indicate the flat film's saturation magnetization was slightly lower than that of the
rough film. For a fairly thick film (80 nm), a slightly reduced saturation
magnetization would be consistent with a thin, magnetically-dead surface layer.
Unfortunately, these researchers only measured conductivity maps at one temperature and
without an applied magnetic field.

\subsection{Causes of the Inactive Layer}

Because the STM tunneling current decreases exponentially with tunneling distance,
\cite{Stroscio1993} only the topmost conducting unit cell of the manganite will
directly contribute to the d$I$/d$V$ spectra.  If the surface unit cell is
electronically representative of the film bulk electronic structure, then the
d$I$/d$V$ spectra will be a measure of film bulk properties. However, if the surface
unit-cell is not electronically representative of the film bulk but still
conducting, then the d$I$/d$V$ spectra will only represent surface layer properties.
The film surface may differ electronically form the film bulk because it is
chemically different~: depleted or enriched in oxygen or dopants, or because the
film crystal structure is necessarily truncated by the film surface.

\subsubsection{Dopant Enrichment and Depletion}

Some studies have found evidence that LCMO films are not chemically homogeneous.
Simon \textit{et al} measured the chemical profile of a LCMO/STO/LCMO trilayer grown
on an STO substrate. \cite{Simon2004} Using energy-loss spectroscopic profiling,
they detected enhanced Ca doping at the surface of the first LCMO layer. They
attributed this to a La-rich layer that forms adjacent to the STO substrate as film
growth initiates. As there is lattice mismatch between LCMO and STO, at the
film-substrate interface the film will be under tensile strain, favoring
substitution of La$^{2+}$, the slightly larger ion (1.36\AA), for Ca$^{3+}$, the
slightly smaller ion (1.34\AA). Because a fixed ratio of LCMO precursors is
sputtered onto the substrate, a Ca-rich layer must form on top of the La-rich layer.
Once formed, the Ca-rich layer floats on the surface while the film grows beneath
it. However, this explanation may not hold; Choi \textit{et al} \cite{Choi1999}
measured a Ca-enriched surface layer on an LCMO film on LAO (where no tensile strain
occurs) using angle-resolved XPS. They found a best-fit model of their data to be a
single Ca-enriched layer at the film surface (60\% Ca doping), significantly above
the nominal 35\% film doping. They hypothesize that this enriched layer was caused
by a large heat of segregation driving Ca towards the surface. Estrad\'{e}
\textit{et al} measured the chemical composition of LCMO films grown on both (001)
and (110) oriented STO substrates with electron energy-loss spectroscopy.
\cite{Estrade2008}They found differences between the surface and bulk concentrations
only for the (001) orientation, although both orientations yield the same strain.
Important for our purpose is that the segregation they found is only small. Although
we have not studied our film's surface chemistry, we find the same inactive layer on
both strained flat films on STO and unstrained flat films on NGO, suggesting that,
if there is any chemical stratification, substrate-induced strain is not the cause.

\subsubsection{Oxygen Depletion}

Oxygen depletion can also shift the Mn valence. In some studies the surface Mn
valence was measured directly and conclusions were drawn on surface oxygen
depletion. Using XPS, Beyreuther \textit{et al} measured exchange splitting of the
Mn 3s core-level peak in LCMO on STO.\cite{Beyreuther2006} From this data they could
infer the Mn valence in the top 3~nm of the film. For as-grown films, the Mn valence
was +3.3, the expected bulk film value, but for films subsequently exposed to UHV at
elevated temperatures, the Mn valence shifted towards +2. However, they found that
the surface Mn valence could be restored to bulk value by reoxidizing the film in
oxygen at elevated temperatures (470$^\circ$C). From this they conclude that the
shifts in Mn valence were driven by oxygen depletion and enrichment. Other research
has demonstrated that exposure to only moderate temperatures (80$^\circ$C) in air is
required to reoxidize LCMO films,\cite{Dorr2000} and this suggests that an
oxygen-poor atmosphere---such as helium gas---at similar temperatures could lead to
oxygen depletion. However, many of our flat and rough films were measured in ambient
before exposure to an oxygen-poor environment, and we could detect no difference
between these measurements and subsequent measurements made in helium gas or UHV.
Because we do not measure our films \textit{in-situ}, it is possible that exposure
to ambient conditions could degrade the surface in some way. This possibility has
been investigated by other researchers. Valencia \textit{et al} measured LCMO films
grown on LAO using x-ray absorption spectroscopy (XAS). \cite{Valencia2006} For
as-grown films, they measured spectra congruent with the expected
Mn$^{3+}$/Mn$^{4+}$ ratio, but once the film had been exposed to air, they also
detected the signature of Mn$^{2+}$ ions. They propose the CO present in air as a
reducing agent that attacks the unstable Mn$^{3+}$ and Mn$^{4+}$ ions. However, as
they note, on films exposed to air for less than 2 days, they could not detect
Mn$^{2+}$ ions. Some of our films were measured after less than an hour of exposure
to air, and we saw no difference between these films and those exposed to air for
significantly longer. If the surface of our as-grown films is oxygen depleted, then
this oxygen depletion seems likely to occur during growth or while cooling from the
film-growth temperature.

\subsubsection{Symmetry Breaking}

Theoretical modeling of doped manganites has indicated non bulk-like properties at the
surface. Calculations by Calderon \textit{et al} for the MnO$_2$-terminated surface of
La$_{1-x}$A$_{x}$MnO$_{3}$ with $x<0.5$, suggest that truncation of the crystal by the
surface leads to charge transfer from the bulk to the surface Mn
layer.\cite{Calderon1999} This charge transfer will create a Mn$^{3+}$ ion-only layer,
suppressing the double exchange interaction, and leading to an anti-ferromagnetic (AFM)
coupling within this layer. These calculations also demonstrate that transport
perpendicular to the surface layer is significantly reduced for an AFM-coupled surface
when compared to a ferromagnetically-coupled surface. Filippetti \textit{et al}
considered ferromagnetic (FM) versus AFM coupling between the surface and immediate
subsurface layer of MnO$_2$-terminated
La$_{1x}$Ca$_{x}$MnO$_{3}$.\cite{Filippetti1999,Filippetti2000} They found the lowest
energy configuration to be FM coupling, regardless of the underlying magnetic order.
However, all of the theoretical modeling of manganite surfaces to date has considered
only planar, singly-terminated surfaces. Our flat samples are planar and, if singly
terminated, most likely terminated by an MnO$_2$ layer.\cite{Choi1999,Akhtar2006} For a
flat film terminated with such an MnO$_2$ layer, the modeling by Calderon \textit{et
al}\cite{Calderon1999} would explain why our flat films are inactive. These results are
less relevant for rough LCMO films which are certainly not planar, and likely not
singly-terminated.

\subsubsection{Further Remarks}

The effect of oxygen depletion or dopant enrichment of the surface layer would
change its properties. Oxygen depletion has been shown to reduce the number of
Mn$^{4+}$ ions in bulk crystal, Ca-doped manganites,\cite{Malavasi2002} and thus, if
the surface were oxygen-depleted, the surface layer would have the properties of  a
manganite with lower hole doping, possibly with a ferromagnetic-insulating ground
state.\cite{Schiffer1995} Similarly, surface enrichment to 60\% Ca doping (as found
by Choi \textit{et al} \cite{Choi1999}) would render the film surface layer ground
state an antiferromagnetic insulator.\cite{Schiffer1995} Whether the surface layer
is a ferromagnetic insulator or an antiferromagnetic insulator below the transition,
measured d$I$/d$V$ spectra at temperatures above and below the transition would
probably only differ slightly---both having the characteristic of insulators, with a
gap or pseudo-gap at zero bias. Distinguishing between a transition of this nature
and a completely inactive surface layer with STS could be difficult. For our
as-grown flat films, either an oxygen-depleted or dopant-enriched surface could
produce the inactive behavior, apparent or real, we observed with STS on atomically
flat films. However, we have grown rough films under identical process conditions to
flat films (changing only the growth time), and found the rough films to be active.
We would expect similar oxygen depletion or dopant enrichment to occur in both rough
and flat films (or even larger in the rough films), and, as such, neither can
explain the behavior of both active and inactive films.

As our rough, thicker films are grown under nominally identical process conditions to
our flat, thinner films, it seems that the change of growth mode underlying the rough
morphology or the rough morphology itself must explain the electronic inhomogeneities
seen on rough films. If a change in the growth mode were to introduce chemical
inhomogeneities into the film, then thick films would consist of two layers: a
chemically homogeneous layer adjacent to the substrate, with a chemically inhomogeneous
layer extending from that layer to the film surface. The chemically homogeneous layer
could be A-site ordered, with the inhomogeneous layer being A-site disordered.
Moshnyaga \textit{et al} compared two LCMO films, one flat and one rough, and found
that whereas the flat film was both electronically homogeneous and A-site ordered, the
rough film was neither electronically homogeneous, nor A-site
ordered.\cite{Moshnyaga2006} A-site disorder could provide a mechanism to disrupt the
poorly conducting and apparently magnetically inactive AFM surface layer proposed by
Calderon \textit{et al}.\cite{Calderon1999} However, the rough morphology could itself
explain the active surface of rough films. Our rough samples, even if epitaxial, would
most likely not be singly terminated, and would also have greater crystal symmetry
breaking because of the non-planar character of the surface. Perhaps a combination of
greater symmetry breaking and a multiply-terminated surface explains why rough films
are active. Flat films could be inactive because the surface layer is locked into AFM
coupling, regardless of applied field or temperature.

\subsection{Attempts to Eliminate the Inactive Layer}

Our attempts to eliminate the inactive surface layer either by annealing or etching
failed to recover the expected bulk film behavior. Oxygen annealing resulted in
modifying the surface topography, rendering a direct comparison to our as-grown films
suspect. Further, exposing samples to elevated temperatures is likely to have resulted
in film relaxation, and introduced defects, especially for strained films on STO. Thus
even the film bulk would no longer be as-grown. Plasma etching necessarily modified the
surface topography, and possibly changed the surface chemistry by preferential
sputtering. Our data from etched films, with metallic $I$-$V$ curves measured in a
narrow temperature window just below $T_{p}$, do agree with $R$-$T$ measurements made
on oxygen-depleted LCMO films by D\"{o}rr \textit{et al}.\cite{Dorr2000} On their most
oxygen-depleted film, the $R$-$T$ curve reaches a peak at the ferromagnetic ordering
temperature before decreasing as the temperature is reduced---as expected for a
paramagnetic-insulator to ferromagnetic-metal transition. But as the temperature is
reduced further, the film resistance again increases, becoming semiconductor-like, with
thermally activated behavior. They argue that oxygen depletion reduces the proportion
of Mn$^{4+}$ ions, reducing the effective doping. As they note, at this doping level,
theoretical modeling by Yunoki \textit{et al}\cite{Yunoki1998} predicted phase
separation of FM and AFM phases at low temperatures, with conduction depending upon
percolation between FM regions. For our etched samples, if the etching had
preferentially removed calcium or oxygen from the surface, the effective hole-doping
would also be reduced. Thus the surface---as measured by STS---could follow a similar
pattern. At high temperatures the film would be paramagnetic and semiconductor-like. As
the temperature is lowered, a transition to a metallic FM phase would take place. Upon
further temperature reduction, the film would enter a phase-separated FM-AFM phase.
This could explain the temperature window of metallic $I$-$V$ curves measured on our
etched films.

%%%%%%%%%%%%%%%%%%%%%%%%%%%%%%%%%%%%%%%%%%%%%%%%%%%%%%%%%%%%%%%%%%%%%%%%%%%%%%%%%%%%%%
\section{Conclusion}
%%%%%%%%%%%%%%%%%%%%%%%%%%%%%%%%%%%%%%%%%%%%%%%%%%%%%%%%%%%%%%%%%%%%%%%%%%%%%%%%%%%%%%

We have used d$I$/d$V$ spectra calculated from STS $I$-$V$ measurements as a proxy for
the sample DOS. With this method, we measured strained and unstrained LCMO films, and
LCMO films with both flat and rough morphology. We detected inactive surface layers on
all of our flat films. Transport measurements confirm that, although the surface layer
of these films are inactive, the film bulk is active---having a $M$-$I$ transition and
being magnetoresistive. On thicker films, whether grown on STO or NGO, the surface
morphology roughens during growth, and is subsequently electronically inhomogeneous. We
find that the surface of these as-grown rough films responds to an externally applied
magnetic field by becoming more metallic, as would be expected if percolation of
metallic regions were to underlie CMR. On flat samples, we attempted to remove the
inactive layer using ion etching. However, the post-etch d$I$/d$V$ spectra were not as
expected and became more insulator-like at low temperatures. Transport measurements
indicate that for a stochiometric film the resistance should decrease at low
temperatures. Etching is likely to have changed the stochiometry by preferentially
etching one or more of the film's chemical species. We oxygen-annealed other flat films
to rule out oxygen deficiency as causing of the inactive layer. Unfortunately,
annealing roughened the surface, rendering a direct comparison with pre-annealing flat
samples questionable. STS and conductivity maps of these roughened films showed them to
be electronically inhomogeneous. Strain does not seem to play a direct role in the
formation of the inactive surface layer as results from LCMO films grown on NGO and STO
were very similar.

Summarizing our quite large set of data, we do find that surface morphology is
critical to our results. Rough films, whether as-grown or roughened by annealing,
were active and electronically inhomogeneous, basically confirming a number of
earlier results. Flat films were inactive and homogeneous. Taking data from
different samples together, we did not find conclusive evidence for the presence of
polaron peaks, nor for a formation of a pseudogap in the $M$-$I$ transition.
Instead, we come to the conclusion that the surface layer of flat films is
insulating, not from chemical causes, but due to the symmetry breaking at the
surface. In line with this we put forward that the film roughness itself, regardless
of its origin, removes this homogeneous symmetry breaking and creates a
multiply-terminated surface that is not locked into an inert magnetic order.

\begin{acknowledgments}
We acknowledge C. Beekman, R. Hendikx and M. Hesselbert for discussions and
technical support. This research is in part supported by NanoNed, a national
nanotechnology program coordinated by the Dutch Ministry of Economic Affairs and in
part by a research program of the Stichting "F.O.M.," which is financially supported
by the Dutch national science foundation NWO.
\end{acknowledgments}

%%%%%%%%%%%%%%%%%%%%%%%%%%%%%%%%%%%%%%%%%%%%%%%%%%%%%%%%%%%%%%%%%%%%%%%%%%%%%%%%%%%%%%
\appendix* \section{}
%%%%%%%%%%%%%%%%%%%%%%%%%%%%%%%%%%%%%%%%%%%%%%%%%%%%%%%%%%%%%%%%%%%%%%%%%%%%%%%%%%%%%%

Consider tunneling between two metals electrodes separated by a vacuum gap. Ignoring
any image potential, and assuming a flat DOS in the vicinity of the Fermi Energy,
E$_{f}$, for both materials, a number of researchers\cite{Ukraintsev1995,
Simmons1963} have shown that the tunneling current is given by \begin{eqnarray}
\label{tunnelingCurrent} I(s,V)= \frac{2A }{ B^{2}}\bigg\{e^{
-B(\bar{\Phi}-\frac{eV}{2})}\bigg[1+B\Big(\bar{\Phi}-\frac{eV}{2}\Big) \bigg]
\nonumber\\ - e^{
-B(\bar{\Phi}+\frac{eV}{2})}\bigg[1+B\Big(\bar{\Phi}+\frac{eV}{2}\Big) \bigg]\bigg\},
\end{eqnarray} where $\bar{\Phi}$ is the average of the two electrode's work
functions, $V_{bias}$ is the bias voltage applied between the electrodes, $s$ is the
distance between the electrodes, $A$ is electrode area, and $B=2s(2m/
\hbar^{2})^{1/2}$. Applying the same assumptions as used in Equation
\ref{tunnelingCurrent}, tunneling conductance can be written as \begin{equation}
\label{tunnelingConductance} \sigma(s,V)= \frac{eA }{2}\bigg\{e^{
-B(\bar{\Phi}-\frac{eV}{2})} + e^{ -B(\bar{\Phi}+\frac{eV}{2})}\bigg\}. \end{equation}
This model can also be used to approximate STS, though the value of $A$ is difficult
to measure or calculate for an STM tip. However, if $s$ does not change significantly
for different bias and current set-points, normalizing will eliminate the need to
know $A$ specifically.

%+++++++++++++++++++++++++++++++++++++++++++++++++++++++++++++++++++++++++++++++++++++
\begin{figure} \centering

\epsfig{file=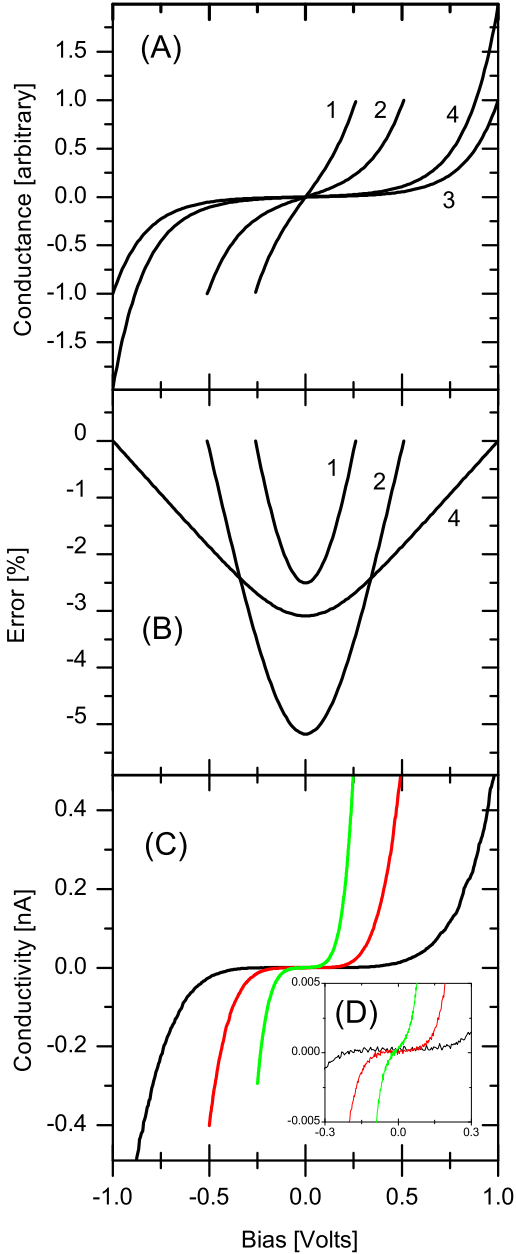,width=8.1cm,clip=}

\caption{\label{calculatedIV}(Color online) (A) Calculated $I$-$V$ curves.
$I_{set}$=1.0 at $V_{set }$=1.0 (1), 0.5 (2), and 0.25V (3); and $I_{set}$=2.0 at
$V_{set}$=1.0V (4). Normalized to tunneling current at 1.0V, and 5nm. (B) Percentage
error of scaled calculated $I$-$V$ curves. (C) Measured $I$-$V$ curves. Same data as
Fig. \ref{STOdIdV} at $T$=155K. $I_{set}$=0.5nA. $V_{set}$=1.0V, $\pm$1.0V (black);
$V_{set}$=0.5V, $\pm$0.5V (red); and $V_{set}$=0.25V, $\pm$0.25V (green). (D) Data
from (c) with enlarged scale.}

\end{figure}
%+++++++++++++++++++++++++++++++++++++++++++++++++++++++++++++++++++++++++++++++++++++

Using Equation \ref{tunnelingCurrent} and using the average of the work function of
LCMO ($\sim4.8 eV$) and Pt ($\sim5.3 eV$),\cite{Shang2006} the current for $I$-$V$
curves can be calculated. This is shown in Figure \ref{calculatedIV}, panel (A).
Three curves show $I_{set}=1.0$ at $V_{set}=1.0$, $0.5$, and $0.25$ volts, and one
curve shows show $I_{set}=2.0$ at $V_{set}=1.0$ volts (curves 1,2,3, and 4
respectively). Curves 1, 2, and 4 were generated by solving numerically for the value
of tunneling gap distance that generated the desired tunneling current. Those values
were $4.81$, $4.87$, and $4.97$ nm respectively. All curves are normalized to the
tunneling current at $V_{bias}=1.0$ volt and $s=5.0$ nm.

Although Equation \ref{tunnelingCurrent} is non-linear, calculated $I$-$V$ curves can
be approximately scaled to each other. Curves 1, 2, and 4 were scaled so as to match
their tunneling current at $V_{set}$ to the tunneling current of curve 3 at the same
voltage. Shown in panel (B) of Figure \ref{calculatedIV} is the percentage error for
the three scaled curves when compared to curve 3. For bias values less than their own
$V_{set}$, the maximum error is 5.2\% at $V_{bias}=0.0$ for curve 2. Calculations
were also performed at higher $I_{set}$ (or equivalently, smaller $s$), and the
maximum error was found to be reduced.

\begin{table} \caption{\label{conductance} Normalized conductance for $I$-$V$ curves
shown in \ref{calculatedIV}, panel (A).} \begin{tabular}{|l|c|c|c|c|} \hline Curve &
1 & 2 & 3 & 4\\ \hline $V=V_{set}$ & 1.1 & 0.96 & 1.0 & 2.0\\ \hline $V=0.0$ & 86.5 &
19.9 & 1.0 & 2.1\\ \hline \end{tabular}\end{table}

To understand the value of scaling low $V_{set}$ $I$-$V$ curves, consider the
tunneling conductance at low bias values. The tunneling conductance at $V_{bias}=0.0$
is just the slope of the $I$-$V$ curve at that bias. Evident from the $I$-$V$ curves
plotted in panel (A) of Figure \ref{calculatedIV}, the tunneling conductance for low
$V_{set}$ curves is much higher. Tunneling conductance can be calculated using
Equation \ref{tunnelingConductance}, and the result of such calculations are shown in
table \ref{conductance}. Both conductance at $V_{set}$ and $V_{bias}=0.0$ are shown.
All data are normalized to the corresponding conductance of curve 3. Note that curve
4, where $I_{set}$ is double that of curve 3, the zero-bias conductance is also only
double that of curve 3. In contrast, whereas $I_{set}$ is identical for curves 1, 2,
and 3, the zero-bias conductance of curve 1 and 2 are, respectively, 86.5 and 19.9
times larger. This difference could be critical for materials with a depleted DOS
around E$_{f}$.

To measure the DOS across a broad energy range, either $V_{set}$ must be increased,
or the bias must be swept over a larger voltage range. As has been demonstrated,
increasing $V_{set}$ will reduce the low-bias conductance, to perhaps below the STM
noise level. The alternative, increasing the sweep range, also faces problems. Most
current-voltage amplifiers have limited dynamic range, and increasing the sweep
range could result in saturating the amplifier. To illustrate this, consider setting
$V_{set}=0.25$ volts, but sweeping $V_{bias}=\pm1.0$ volt. Using Equation
\ref{tunnelingCurrent} and the tunneling gap used for curve 1, we find the
conductivity at $V_{bias}=1.0$ to be approximately 70 times larger than at
$V_{set}=0.25$. Other solutions include variable-gap STS or fixed-gap STS with a
non-linear current-voltage amplifier. Variable-gap STS requires very accurate
calibration of the STM tip actuator (usually a piezo tube) due to the exponential
dependence of tunneling current on tip-sample distance.

In the above numerical examples, a flat DOS was assumed. Nevertheless, though we do
not expect our samples to have a flat DOS, particularly at E$_{f}$, we were able to
successfully scale our data using this method. Panel (C) of Figure \ref{calculatedIV}
shows unprocessed $I$-$V$ curves for the 155K data shown in Figure \ref{STOdIdV}.
These unprocessed $I$-$V$ curves were scaled as described above, and then numerically
differentiated. No further adjustment was made after differentiation. As can be seen
in Figure \ref{STOdIdV}, the adjusted curves align well with each other, demonstrating
the effectiveness of this method, even for samples with a depleted DOS at zero bias.
Note that the $V_{bias}=\pm1.0$ d$I$/d$V$ spectra in Figure \ref{STOdIdV} is very
noisy between $V_{bias}=\pm0.3$ whereas both the $V_{bias}=\pm0.25$ and $\pm0.5$
d$I$/d$V$ spectra are much less noisy over much of this range. Panel (D) of Figure
\ref{calculatedIV} shows data from panel (C) with an enlarged scale to illustrate the
greatly increased zero-bias conductance for smaller $V_{set}$ $I$-$V$ measurements.
The differences between these three curves are almost entirely due to changes in the
tunneling barrier caused by changing $V_{set}$.

%%%%%%%%%%%%%%%%%%%%%%%%%%%%%%%%%%%%%%%%%%%%%%%%%%%%%%%%%%%%%%%%%%%%%%%%%%%%%%%%%%%%%%
%%%%%%%%%%%%%%%%%%%%%%%%%%%%%%%%%%%%%%%%%%%%%%%%%%%%%%%%%%%%%%%%%%%%%%%%%%%%%%%%%%%%%%
%% THE END
%%%%%%%%%%%%%%%%%%%%%%%%%%%%%%%%%%%%%%%%%%%%%%%%%%%%%%%%%%%%%%%%%%%%%%%%%%%%%%%%%%%%%%
%%%%%%%%%%%%%%%%%%%%%%%%%%%%%%%%%%%%%%%%%%%%%%%%%%%%%%%%%%%%%%%%%%%%%%%%%%%%%%%%%%%%%%

\bibliographystyle{apsrev}
\bibliography{P:/Users/Docs/Science/Manuscr/Simon-STM-on-Mnites/STM-Manganites_30jan09/Deadlayer02}

\begin{thebibliography}{41}
\expandafter\ifx\csname natexlab\endcsname\relax\def\natexlab#1{#1}\fi
\expandafter\ifx\csname bibnamefont\endcsname\relax
  \def\bibnamefont#1{#1}\fi
\expandafter\ifx\csname bibfnamefont\endcsname\relax
  \def\bibfnamefont#1{#1}\fi
\expandafter\ifx\csname citenamefont\endcsname\relax
  \def\citenamefont#1{#1}\fi
\expandafter\ifx\csname url\endcsname\relax
  \def\url#1{\texttt{#1}}\fi
\expandafter\ifx\csname urlprefix\endcsname\relax\def\urlprefix{URL }\fi
\providecommand{\bibinfo}[2]{#2}
\providecommand{\eprint}[2][]{\url{#2}}

\bibitem[{\citenamefont{F\"ath et~al.}(1999)\citenamefont{F\"ath, Freisem,
  Menovsky, Tomioka, Aarts, and Mydosh}}]{Fath1999}
\bibinfo{author}{\bibfnamefont{M.}~\bibnamefont{F\"ath}},
  \bibinfo{author}{\bibfnamefont{S.}~\bibnamefont{Freisem}},
  \bibinfo{author}{\bibfnamefont{A.~A.} \bibnamefont{Menovsky}},
  \bibinfo{author}{\bibfnamefont{Y.}~\bibnamefont{Tomioka}},
  \bibinfo{author}{\bibfnamefont{J.}~\bibnamefont{Aarts}}, \bibnamefont{and}
  \bibinfo{author}{\bibfnamefont{J.~A.} \bibnamefont{Mydosh}},
  \bibinfo{journal}{Science} \textbf{\bibinfo{volume}{285}},
  \bibinfo{pages}{1540} (\bibinfo{year}{1999}).

\bibitem[{\citenamefont{Becker et~al.}(2002)\citenamefont{Becker, Streng, Luo,
  Moshnyaga, Damaschke, Shannon, and Samwer}}]{Becker2002}
\bibinfo{author}{\bibfnamefont{T.}~\bibnamefont{Becker}},
  \bibinfo{author}{\bibfnamefont{C.}~\bibnamefont{Streng}},
  \bibinfo{author}{\bibfnamefont{Y.}~\bibnamefont{Luo}},
  \bibinfo{author}{\bibfnamefont{V.}~\bibnamefont{Moshnyaga}},
  \bibinfo{author}{\bibfnamefont{B.}~\bibnamefont{Damaschke}},
  \bibinfo{author}{\bibfnamefont{N.}~\bibnamefont{Shannon}}, \bibnamefont{and}
  \bibinfo{author}{\bibfnamefont{K.}~\bibnamefont{Samwer}},
  \bibinfo{journal}{Phys. Rev. Lett.} \textbf{\bibinfo{volume}{89}},
  \bibinfo{pages}{237203} (\bibinfo{year}{2002}).

\bibitem[{\citenamefont{Chen et~al.}(2003)\citenamefont{Chen, Lin, Juang, Uen,
  Wu, Gou, and Lin}}]{Chen2003}
\bibinfo{author}{\bibfnamefont{S.~F.} \bibnamefont{Chen}},
  \bibinfo{author}{\bibfnamefont{P.~I.} \bibnamefont{Lin}},
  \bibinfo{author}{\bibfnamefont{J.~Y.} \bibnamefont{Juang}},
  \bibinfo{author}{\bibfnamefont{T.~M.} \bibnamefont{Uen}},
  \bibinfo{author}{\bibfnamefont{K.~H.} \bibnamefont{Wu}},
  \bibinfo{author}{\bibfnamefont{Y.~S.} \bibnamefont{Gou}}, \bibnamefont{and}
  \bibinfo{author}{\bibfnamefont{J.~Y.} \bibnamefont{Lin}},
  \bibinfo{journal}{Appl. Phys. Lett.} \textbf{\bibinfo{volume}{82}},
  \bibinfo{pages}{1242} (\bibinfo{year}{2003}).

\bibitem[{\citenamefont{Seiro et~al.}(2008)\citenamefont{Seiro, Fasano,
  Maggio-Aprile, Koller, Kuffer, and Fischer}}]{Seiro2008}
\bibinfo{author}{\bibfnamefont{S.}~\bibnamefont{Seiro}},
  \bibinfo{author}{\bibfnamefont{Y.}~\bibnamefont{Fasano}},
  \bibinfo{author}{\bibfnamefont{I.}~\bibnamefont{Maggio-Aprile}},
  \bibinfo{author}{\bibfnamefont{E.}~\bibnamefont{Koller}},
  \bibinfo{author}{\bibfnamefont{O.}~\bibnamefont{Kuffer}}, \bibnamefont{and}
  \bibinfo{author}{\bibfnamefont{.}~\bibnamefont{Fischer}},
  \bibinfo{journal}{Phys. Rev. B} \textbf{\bibinfo{volume}{77}},
  \bibinfo{pages}{020407(R)} (\bibinfo{year}{2008}).

\bibitem[{\citenamefont{Biswas et~al.}(2001)\citenamefont{Biswas, Rajeswari,
  Srivastava, Venkatesan, Greene, Lu, de~Lozanne, and Millis}}]{Biswas2001}
\bibinfo{author}{\bibfnamefont{A.}~\bibnamefont{Biswas}},
  \bibinfo{author}{\bibfnamefont{M.}~\bibnamefont{Rajeswari}},
  \bibinfo{author}{\bibfnamefont{R.~C.} \bibnamefont{Srivastava}},
  \bibinfo{author}{\bibfnamefont{T.}~\bibnamefont{Venkatesan}},
  \bibinfo{author}{\bibfnamefont{R.~L.} \bibnamefont{Greene}},
  \bibinfo{author}{\bibfnamefont{Q.}~\bibnamefont{Lu}},
  \bibinfo{author}{\bibfnamefont{A.~L.} \bibnamefont{de~Lozanne}},
  \bibnamefont{and} \bibinfo{author}{\bibfnamefont{A.~J.}
  \bibnamefont{Millis}}, \bibinfo{journal}{Phys. Rev. B}
  \textbf{\bibinfo{volume}{63}}, \bibinfo{pages}{184424}
  (\bibinfo{year}{2001}).

\bibitem[{\citenamefont{Zhang et~al.}(2002)\citenamefont{Zhang, Israel, Biswas,
  Greene, and de~Lozanne}}]{Zhang2002}
\bibinfo{author}{\bibfnamefont{L.}~\bibnamefont{Zhang}},
  \bibinfo{author}{\bibfnamefont{C.}~\bibnamefont{Israel}},
  \bibinfo{author}{\bibfnamefont{A.}~\bibnamefont{Biswas}},
  \bibinfo{author}{\bibfnamefont{R.~L.} \bibnamefont{Greene}},
  \bibnamefont{and}
  \bibinfo{author}{\bibfnamefont{A.}~\bibnamefont{de~Lozanne}},
  \bibinfo{journal}{Science} \textbf{\bibinfo{volume}{298}},
  \bibinfo{pages}{805} (\bibinfo{year}{2002}).

\bibitem[{\citenamefont{Moreo et~al.}(2000)\citenamefont{Moreo, Mayr, Feiguin,
  Yunoki, and Dagotto}}]{Moreo2000}
\bibinfo{author}{\bibfnamefont{A.}~\bibnamefont{Moreo}},
  \bibinfo{author}{\bibfnamefont{M.}~\bibnamefont{Mayr}},
  \bibinfo{author}{\bibfnamefont{A.}~\bibnamefont{Feiguin}},
  \bibinfo{author}{\bibfnamefont{S.}~\bibnamefont{Yunoki}}, \bibnamefont{and}
  \bibinfo{author}{\bibfnamefont{E.}~\bibnamefont{Dagotto}},
  \bibinfo{journal}{Phys. Rev. Lett.} \textbf{\bibinfo{volume}{84}},
  \bibinfo{pages}{5568} (\bibinfo{year}{2000}).

\bibitem[{\citenamefont{Dagotto et~al.}(2008)\citenamefont{Dagotto, Yunoki,
  Sen, Alvarez, and Moreo}}]{Dagotto2008}
\bibinfo{author}{\bibfnamefont{E.}~\bibnamefont{Dagotto}},
  \bibinfo{author}{\bibfnamefont{S.}~\bibnamefont{Yunoki}},
  \bibinfo{author}{\bibfnamefont{C.}~\bibnamefont{Sen}},
  \bibinfo{author}{\bibfnamefont{G.}~\bibnamefont{Alvarez}}, \bibnamefont{and}
  \bibinfo{author}{\bibfnamefont{A.}~\bibnamefont{Moreo}}, \bibinfo{journal}{J.
  Phys.: Condens. Matter} \textbf{\bibinfo{volume}{20}},
  \bibinfo{pages}{434224} (\bibinfo{year}{2008}).

\bibitem[{\citenamefont{Borges et~al.}(2001)\citenamefont{Borges, Guichard,
  Lunney, Coey, and Ott}}]{Borges2001}
\bibinfo{author}{\bibfnamefont{R.~P.} \bibnamefont{Borges}},
  \bibinfo{author}{\bibfnamefont{W.}~\bibnamefont{Guichard}},
  \bibinfo{author}{\bibfnamefont{J.~G.} \bibnamefont{Lunney}},
  \bibinfo{author}{\bibfnamefont{J.~M.~D.} \bibnamefont{Coey}},
  \bibnamefont{and} \bibinfo{author}{\bibfnamefont{F.}~\bibnamefont{Ott}},
  \bibinfo{journal}{J. Appl. Phys.} \textbf{\bibinfo{volume}{89}},
  \bibinfo{pages}{3868} (\bibinfo{year}{2001}).

\bibitem[{\citenamefont{Abad et~al.}(2005)\citenamefont{Abad, Martínez, and
  Balcellsa}}]{Abad2005}
\bibinfo{author}{\bibfnamefont{L.}~\bibnamefont{Abad}},
  \bibinfo{author}{\bibfnamefont{B.}~\bibnamefont{Martínez}}, \bibnamefont{and}
  \bibinfo{author}{\bibfnamefont{L.}~\bibnamefont{Balcellsa}},
  \bibinfo{journal}{Appl. Phys. Lett.} \textbf{\bibinfo{volume}{87}},
  \bibinfo{pages}{212502} (\bibinfo{year}{2005}).

\bibitem[{\citenamefont{Yao et~al.}(2006)\citenamefont{Yao, Zhang, Cui, Wang,
  and Shen}}]{Yao2006}
\bibinfo{author}{\bibfnamefont{Z.}~\bibnamefont{Yao}},
  \bibinfo{author}{\bibfnamefont{L.}~\bibnamefont{Zhang}},
  \bibinfo{author}{\bibfnamefont{Y.}~\bibnamefont{Cui}},
  \bibinfo{author}{\bibfnamefont{C.}~\bibnamefont{Wang}}, \bibnamefont{and}
  \bibinfo{author}{\bibfnamefont{Z.}~\bibnamefont{Shen}},
  \bibinfo{journal}{Solid State Communications} \textbf{\bibinfo{volume}{139}},
  \bibinfo{pages}{465} (\bibinfo{year}{2006}).

\bibitem[{\citenamefont{Valencia et~al.}(2007)\citenamefont{Valencia, Gaupp,
  Gudat, Abad, Balcells, and Martínez}}]{Valencia2007a}
\bibinfo{author}{\bibfnamefont{S.}~\bibnamefont{Valencia}},
  \bibinfo{author}{\bibfnamefont{A.}~\bibnamefont{Gaupp}},
  \bibinfo{author}{\bibfnamefont{W.}~\bibnamefont{Gudat}},
  \bibinfo{author}{\bibfnamefont{L.}~\bibnamefont{Abad}},
  \bibinfo{author}{\bibfnamefont{L.}~\bibnamefont{Balcells}}, \bibnamefont{and}
  \bibinfo{author}{\bibfnamefont{B.}~\bibnamefont{Martínez}},
  \bibinfo{journal}{Appl. Phys. Lett.} \textbf{\bibinfo{volume}{90}},
  \bibinfo{pages}{252509} (\bibinfo{year}{2007}).

\bibitem[{\citenamefont{Choi et~al.}(1999)\citenamefont{Choi, Zhang, Liou,
  Dowben, and Plummer}}]{Choi1999}
\bibinfo{author}{\bibfnamefont{J.}~\bibnamefont{Choi}},
  \bibinfo{author}{\bibfnamefont{J.}~\bibnamefont{Zhang}},
  \bibinfo{author}{\bibfnamefont{S.-H.} \bibnamefont{Liou}},
  \bibinfo{author}{\bibfnamefont{P.~A.} \bibnamefont{Dowben}},
  \bibnamefont{and} \bibinfo{author}{\bibfnamefont{E.~W.}
  \bibnamefont{Plummer}}, \bibinfo{journal}{Phys. Rev. B}
  \textbf{\bibinfo{volume}{59}}, \bibinfo{pages}{13453} (\bibinfo{year}{1999}).

\bibitem[{\citenamefont{Simon et~al.}(2004)\citenamefont{Simon, Walther, Mader,
  Klein, Reisinger, Alff, and Gross)}}]{Simon2004}
\bibinfo{author}{\bibfnamefont{J.}~\bibnamefont{Simon}},
  \bibinfo{author}{\bibfnamefont{T.}~\bibnamefont{Walther}},
  \bibinfo{author}{\bibfnamefont{W.}~\bibnamefont{Mader}},
  \bibinfo{author}{\bibfnamefont{J.}~\bibnamefont{Klein}},
  \bibinfo{author}{\bibfnamefont{D.}~\bibnamefont{Reisinger}},
  \bibinfo{author}{\bibfnamefont{L.}~\bibnamefont{Alff}}, \bibnamefont{and}
  \bibinfo{author}{\bibfnamefont{R.}~\bibnamefont{Gross)}},
  \bibinfo{journal}{Appl. Phys. Lett.} \textbf{\bibinfo{volume}{84}},
  \bibinfo{pages}{3882} (\bibinfo{year}{2004}).

\bibitem[{\citenamefont{Estrad\'{e} et~al.}(2007)\citenamefont{Estrad\'{e},
  Arbiol, Peir\'{o}, Abad, Laukhin, Balcells, and Mart\'{i}nez}}]{Estrade2007}
\bibinfo{author}{\bibfnamefont{S.}~\bibnamefont{Estrad\'{e}}},
  \bibinfo{author}{\bibfnamefont{J.}~\bibnamefont{Arbiol}},
  \bibinfo{author}{\bibfnamefont{F.}~\bibnamefont{Peir\'{o}}},
  \bibinfo{author}{\bibfnamefont{L.}~\bibnamefont{Abad}},
  \bibinfo{author}{\bibfnamefont{V.}~\bibnamefont{Laukhin}},
  \bibinfo{author}{\bibfnamefont{L.}~\bibnamefont{Balcells}}, \bibnamefont{and}
  \bibinfo{author}{\bibfnamefont{B.}~\bibnamefont{Mart\'{i}nez}},
  \bibinfo{journal}{Appl. Phys. Lett.} \textbf{\bibinfo{volume}{91}},
  \bibinfo{pages}{252503} (\bibinfo{year}{2007}).

\bibitem[{\citenamefont{Calder\'{o}n et~al.}(1999)\citenamefont{Calder\'{o}n,
  Brey, and Guinea}}]{Calderon1999}
\bibinfo{author}{\bibfnamefont{M.~J.} \bibnamefont{Calder\'{o}n}},
  \bibinfo{author}{\bibfnamefont{L.}~\bibnamefont{Brey}}, \bibnamefont{and}
  \bibinfo{author}{\bibfnamefont{F.}~\bibnamefont{Guinea}},
  \bibinfo{journal}{Phys. Rev. B} \textbf{\bibinfo{volume}{60}},
  \bibinfo{pages}{6698} (\bibinfo{year}{1999}).

\bibitem[{\citenamefont{Filippetti and Pickett}(2000)}]{Filippetti2000}
\bibinfo{author}{\bibfnamefont{A.}~\bibnamefont{Filippetti}} \bibnamefont{and}
  \bibinfo{author}{\bibfnamefont{W.~E.} \bibnamefont{Pickett}},
  \bibinfo{journal}{Phys. Rev. B} \textbf{\bibinfo{volume}{62}},
  \bibinfo{pages}{11571} (\bibinfo{year}{2000}).

\bibitem[{\citenamefont{Zenia et~al.}(2005)\citenamefont{Zenia, Gehring,
  Banach, and Temmerman}}]{Zenia2005}
\bibinfo{author}{\bibfnamefont{H.}~\bibnamefont{Zenia}},
  \bibinfo{author}{\bibfnamefont{G.~A.} \bibnamefont{Gehring}},
  \bibinfo{author}{\bibfnamefont{G.}~\bibnamefont{Banach}}, \bibnamefont{and}
  \bibinfo{author}{\bibfnamefont{W.~M.} \bibnamefont{Temmerman}},
  \bibinfo{journal}{Phys. Rev. B} \textbf{\bibinfo{volume}{71}},
  \bibinfo{pages}{024416} (\bibinfo{year}{2005}).

\bibitem[{\citenamefont{Wei et~al.}(1997)\citenamefont{Wei, Yeh, and
  Vasquez}}]{Wei1997}
\bibinfo{author}{\bibfnamefont{J.~Y.~T.} \bibnamefont{Wei}},
  \bibinfo{author}{\bibfnamefont{N.-C.} \bibnamefont{Yeh}}, \bibnamefont{and}
  \bibinfo{author}{\bibfnamefont{R.~P.} \bibnamefont{Vasquez}},
  \bibinfo{journal}{Phys. Rev. Lett.} \textbf{\bibinfo{volume}{79}},
  \bibinfo{pages}{5150} (\bibinfo{year}{1997}).

\bibitem[{\citenamefont{Mitra et~al.}(2005)\citenamefont{Mitra, Paranjape,
  Raychaudhuri, Mathur, and Blamire}}]{Mitra2005}
\bibinfo{author}{\bibfnamefont{J.}~\bibnamefont{Mitra}},
  \bibinfo{author}{\bibfnamefont{M.}~\bibnamefont{Paranjape}},
  \bibinfo{author}{\bibfnamefont{A.~K.} \bibnamefont{Raychaudhuri}},
  \bibinfo{author}{\bibfnamefont{N.~D.} \bibnamefont{Mathur}},
  \bibnamefont{and} \bibinfo{author}{\bibfnamefont{M.~G.}
  \bibnamefont{Blamire}}, \bibinfo{journal}{Phys. Rev. B}
  \textbf{\bibinfo{volume}{71}}, \bibinfo{pages}{94426} (\bibinfo{year}{2005}).

\bibitem[{\citenamefont{Beekman et~al.}(2007)\citenamefont{Beekman, Komissarov,
  Hesselberth, and Aarts}}]{Beekman2007}
\bibinfo{author}{\bibfnamefont{C.}~\bibnamefont{Beekman}},
  \bibinfo{author}{\bibfnamefont{I.}~\bibnamefont{Komissarov}},
  \bibinfo{author}{\bibfnamefont{M.}~\bibnamefont{Hesselberth}},
  \bibnamefont{and} \bibinfo{author}{\bibfnamefont{J.}~\bibnamefont{Aarts}},
  \bibinfo{journal}{Appl. Phys. Lett.} \textbf{\bibinfo{volume}{91}},
  \bibinfo{pages}{62101} (\bibinfo{year}{2007}).

\bibitem[{\citenamefont{Wittneven et~al.}(1997)\citenamefont{Wittneven,
  Dombrowski, Pan, and Wiesendanger}}]{Wittneven1997}
\bibinfo{author}{\bibfnamefont{C.}~\bibnamefont{Wittneven}},
  \bibinfo{author}{\bibfnamefont{R.}~\bibnamefont{Dombrowski}},
  \bibinfo{author}{\bibfnamefont{S.}~\bibnamefont{Pan}}, \bibnamefont{and}
  \bibinfo{author}{\bibfnamefont{R.}~\bibnamefont{Wiesendanger}},
  \bibinfo{journal}{Rev. Sci. Instrum.} \textbf{\bibinfo{volume}{68}},
  \bibinfo{pages}{3806} (\bibinfo{year}{1997}).

\bibitem[{\citenamefont{Sanchez et~al.}(2006)\citenamefont{Sanchez, Infante,
  Luders, Abad, and Fontcuberta}}]{Sanchez2006}
\bibinfo{author}{\bibfnamefont{F.}~\bibnamefont{Sanchez}},
  \bibinfo{author}{\bibfnamefont{I.}~\bibnamefont{Infante}},
  \bibinfo{author}{\bibfnamefont{U.}~\bibnamefont{Luders}},
  \bibinfo{author}{\bibfnamefont{L.}~\bibnamefont{Abad}}, \bibnamefont{and}
  \bibinfo{author}{\bibfnamefont{J.}~\bibnamefont{Fontcuberta}},
  \bibinfo{journal}{Surface Science} \textbf{\bibinfo{volume}{600}},
  \bibinfo{pages}{1231} (\bibinfo{year}{2006}).

\bibitem[{\citenamefont{Beyreuther et~al.}(2006)\citenamefont{Beyreuther,
  Grafström, Eng, Thiele, and D\"{o}rr}}]{Beyreuther2006}
\bibinfo{author}{\bibfnamefont{E.}~\bibnamefont{Beyreuther}},
  \bibinfo{author}{\bibfnamefont{S.}~\bibnamefont{Grafström}},
  \bibinfo{author}{\bibfnamefont{L.~M.} \bibnamefont{Eng}},
  \bibinfo{author}{\bibfnamefont{C.}~\bibnamefont{Thiele}}, \bibnamefont{and}
  \bibinfo{author}{\bibfnamefont{K.}~\bibnamefont{D\"{o}rr}},
  \bibinfo{journal}{Phys. Rev. B} \textbf{\bibinfo{volume}{73}},
  \bibinfo{pages}{155425} (\bibinfo{year}{2006}).

\bibitem[{\citenamefont{Stroscio and Feenstra}(1993)}]{Stroscio1993}
\bibinfo{author}{\bibfnamefont{J.~A.} \bibnamefont{Stroscio}} \bibnamefont{and}
  \bibinfo{author}{\bibfnamefont{R.}~\bibnamefont{Feenstra}}, in
  \emph{\bibinfo{booktitle}{Scanning Tunneling Microscopy}}, edited by
  \bibinfo{editor}{\bibfnamefont{J.}~\bibnamefont{Stroscio}} \bibnamefont{and}
  \bibinfo{editor}{\bibfnamefont{W.}~\bibnamefont{Kaiser}}
  (\bibinfo{publisher}{Academic Press Inc.}, \bibinfo{year}{1993}),
  vol.~\bibinfo{volume}{27} of \emph{\bibinfo{series}{Methods of Experimental
  Physics}}, chap.~\bibinfo{chapter}{4}, p.~\bibinfo{pages}{95}.

\bibitem[{\citenamefont{Hudson et~al.}(2001)\citenamefont{Hudson, Cohen, Yates,
  Damay, MacManus-Driscoll, Mathur, Blamire, Pakes, and
  Josephs-Franks}}]{Hudson2001}
\bibinfo{author}{\bibfnamefont{P.}~\bibnamefont{Hudson}},
  \bibinfo{author}{\bibfnamefont{L.}~\bibnamefont{Cohen}},
  \bibinfo{author}{\bibfnamefont{K.}~\bibnamefont{Yates}},
  \bibinfo{author}{\bibfnamefont{F.}~\bibnamefont{Damay}},
  \bibinfo{author}{\bibfnamefont{J.}~\bibnamefont{MacManus-Driscoll}},
  \bibinfo{author}{\bibfnamefont{N.}~\bibnamefont{Mathur}},
  \bibinfo{author}{\bibfnamefont{M.}~\bibnamefont{Blamire}},
  \bibinfo{author}{\bibfnamefont{C.}~\bibnamefont{Pakes}}, \bibnamefont{and}
  \bibinfo{author}{\bibfnamefont{P.}~\bibnamefont{Josephs-Franks}},
  \bibinfo{journal}{J. Magn. Magn. Mater.} \textbf{\bibinfo{volume}{226-230}},
  \bibinfo{pages}{2007} (\bibinfo{year}{2001}).

\bibitem[{\citenamefont{Biswas et~al.}(1999)\citenamefont{Biswas, Elizabeth,
  Raychaudhuri, and Bhat}}]{Biswas1999}
\bibinfo{author}{\bibfnamefont{A.}~\bibnamefont{Biswas}},
  \bibinfo{author}{\bibfnamefont{S.}~\bibnamefont{Elizabeth}},
  \bibinfo{author}{\bibfnamefont{A.~K.} \bibnamefont{Raychaudhuri}},
  \bibnamefont{and} \bibinfo{author}{\bibfnamefont{H.~L.} \bibnamefont{Bhat}},
  \bibinfo{journal}{Phys. Rev. B} \textbf{\bibinfo{volume}{59}},
  \bibinfo{pages}{5368} (\bibinfo{year}{1999}).

\bibitem[{\citenamefont{Paranjape et~al.}(2003)\citenamefont{Paranjape,
  Raychaudhuri, Mathur, and Blamire}}]{Paranjape2003}
\bibinfo{author}{\bibfnamefont{M.}~\bibnamefont{Paranjape}},
  \bibinfo{author}{\bibfnamefont{A.~K.} \bibnamefont{Raychaudhuri}},
  \bibinfo{author}{\bibfnamefont{N.~D.} \bibnamefont{Mathur}},
  \bibnamefont{and} \bibinfo{author}{\bibfnamefont{M.~G.}
  \bibnamefont{Blamire}}, \bibinfo{journal}{Phys. Rev. B}
  \textbf{\bibinfo{volume}{67}}, \bibinfo{pages}{214415}
  (\bibinfo{year}{2003}).

\bibitem[{\citenamefont{Ukraintsev}(1995)}]{Ukraintsev1995}
\bibinfo{author}{\bibfnamefont{V.}~\bibnamefont{Ukraintsev}},
  \bibinfo{journal}{Phys. Rev. B} \textbf{\bibinfo{volume}{53}},
  \bibinfo{pages}{11176} (\bibinfo{year}{1995}).

\bibitem[{\citenamefont{Koslowski et~al.}(2007)\citenamefont{Koslowski,
  Dietrich, Tschetschetkin, and Ziemann}}]{Koslowski2007}
\bibinfo{author}{\bibfnamefont{B.}~\bibnamefont{Koslowski}},
  \bibinfo{author}{\bibfnamefont{C.}~\bibnamefont{Dietrich}},
  \bibinfo{author}{\bibfnamefont{A.}~\bibnamefont{Tschetschetkin}},
  \bibnamefont{and} \bibinfo{author}{\bibfnamefont{P.}~\bibnamefont{Ziemann}},
  \bibinfo{journal}{Phys. Rev. B} \textbf{\bibinfo{volume}{75}},
  \bibinfo{pages}{035421} (\bibinfo{year}{2007}).

\bibitem[{\citenamefont{Moshnyaga et~al.}(2006)\citenamefont{Moshnyaga,
  Sudheendra, Lebedev, Koster, Gehrke, Shapoval, Belenchuk, Damaschke, van
  Tendeloo, and Samwer}}]{Moshnyaga2006}
\bibinfo{author}{\bibfnamefont{V.}~\bibnamefont{Moshnyaga}},
  \bibinfo{author}{\bibfnamefont{L.}~\bibnamefont{Sudheendra}},
  \bibinfo{author}{\bibfnamefont{O.~I.} \bibnamefont{Lebedev}},
  \bibinfo{author}{\bibfnamefont{S.~A.} \bibnamefont{Koster}},
  \bibinfo{author}{\bibfnamefont{K.}~\bibnamefont{Gehrke}},
  \bibinfo{author}{\bibfnamefont{O.}~\bibnamefont{Shapoval}},
  \bibinfo{author}{\bibfnamefont{A.}~\bibnamefont{Belenchuk}},
  \bibinfo{author}{\bibfnamefont{B.}~\bibnamefont{Damaschke}},
  \bibinfo{author}{\bibfnamefont{G.}~\bibnamefont{van Tendeloo}},
  \bibnamefont{and} \bibinfo{author}{\bibfnamefont{K.}~\bibnamefont{Samwer}},
  \bibinfo{journal}{Phys. Rev. Lett.} \textbf{\bibinfo{volume}{97}},
  \bibinfo{pages}{107205} (\bibinfo{year}{2006}).

\bibitem[{\citenamefont{Estrad\'{e} et~al.}(2008)\citenamefont{Estrad\'{e},
  Arbiol, Peir\'{o}, Infante, S\'{a}nchez, Fontcuberta, de~la Pe\~{n}a
  M.~Walls, and Colliex}}]{Estrade2008}
\bibinfo{author}{\bibfnamefont{S.}~\bibnamefont{Estrad\'{e}}},
  \bibinfo{author}{\bibfnamefont{J.}~\bibnamefont{Arbiol}},
  \bibinfo{author}{\bibfnamefont{F.}~\bibnamefont{Peir\'{o}}},
  \bibinfo{author}{\bibfnamefont{I.~C.} \bibnamefont{Infante}},
  \bibinfo{author}{\bibfnamefont{F.}~\bibnamefont{S\'{a}nchez}},
  \bibinfo{author}{\bibfnamefont{J.}~\bibnamefont{Fontcuberta}},
  \bibinfo{author}{\bibfnamefont{F.}~\bibnamefont{de~la Pe\~{n}a M.~Walls}},
  \bibnamefont{and} \bibinfo{author}{\bibfnamefont{C.}~\bibnamefont{Colliex}},
  \bibinfo{journal}{Appl. Phys. Lett.} \textbf{\bibinfo{volume}{93}},
  \bibinfo{pages}{112505} (\bibinfo{year}{2008}).

\bibitem[{\citenamefont{D\"{o}rr et~al.}(2000)\citenamefont{D\"{o}rr, Teresa,
  M\"{u}ller, Eckert, Walter, Vlakhov, Nenkov, and Schultz}}]{Dorr2000}
\bibinfo{author}{\bibfnamefont{K.}~\bibnamefont{D\"{o}rr}},
  \bibinfo{author}{\bibfnamefont{J.~M.~D.} \bibnamefont{Teresa}},
  \bibinfo{author}{\bibfnamefont{K.-H.} \bibnamefont{M\"{u}ller}},
  \bibinfo{author}{\bibfnamefont{D.}~\bibnamefont{Eckert}},
  \bibinfo{author}{\bibfnamefont{T.}~\bibnamefont{Walter}},
  \bibinfo{author}{\bibfnamefont{E.}~\bibnamefont{Vlakhov}},
  \bibinfo{author}{\bibfnamefont{K.}~\bibnamefont{Nenkov}}, \bibnamefont{and}
  \bibinfo{author}{\bibfnamefont{L.}~\bibnamefont{Schultz}},
  \bibinfo{journal}{J. Phys.: Condens. Matter} \textbf{\bibinfo{volume}{12}},
  \bibinfo{pages}{7099} (\bibinfo{year}{2000}).

\bibitem[{\citenamefont{Valencia et~al.}(2006)\citenamefont{Valencia, Gaupp,
  Gudat, Abad, Balcells, Cavallaro, Martínez, and Palomares}}]{Valencia2006}
\bibinfo{author}{\bibfnamefont{S.}~\bibnamefont{Valencia}},
  \bibinfo{author}{\bibfnamefont{A.}~\bibnamefont{Gaupp}},
  \bibinfo{author}{\bibfnamefont{W.}~\bibnamefont{Gudat}},
  \bibinfo{author}{\bibfnamefont{L.}~\bibnamefont{Abad}},
  \bibinfo{author}{\bibfnamefont{L.}~\bibnamefont{Balcells}},
  \bibinfo{author}{\bibfnamefont{A.}~\bibnamefont{Cavallaro}},
  \bibinfo{author}{\bibfnamefont{B.}~\bibnamefont{Martínez}}, \bibnamefont{and}
  \bibinfo{author}{\bibfnamefont{F.~J.} \bibnamefont{Palomares}},
  \bibinfo{journal}{Phys. Rev. B} \textbf{\bibinfo{volume}{73}},
  \bibinfo{pages}{104402} (\bibinfo{year}{2006}).

\bibitem[{\citenamefont{Filippetti and Pickett}(1999)}]{Filippetti1999}
\bibinfo{author}{\bibfnamefont{A.}~\bibnamefont{Filippetti}} \bibnamefont{and}
  \bibinfo{author}{\bibfnamefont{W.~E.} \bibnamefont{Pickett}},
  \bibinfo{journal}{Phys. Rev. Lett.} \textbf{\bibinfo{volume}{83}},
  \bibinfo{pages}{4184} (\bibinfo{year}{1999}).

\bibitem[{\citenamefont{Akhtar et~al.}(2006)\citenamefont{Akhtar, Catlow,
  Slater, Walker, and Woodley}}]{Akhtar2006}
\bibinfo{author}{\bibfnamefont{M.}~\bibnamefont{Akhtar}},
  \bibinfo{author}{\bibfnamefont{C.}~\bibnamefont{Catlow}},
  \bibinfo{author}{\bibfnamefont{B.}~\bibnamefont{Slater}},
  \bibinfo{author}{\bibfnamefont{A.}~\bibnamefont{Walker}}, \bibnamefont{and}
  \bibinfo{author}{\bibfnamefont{S.}~\bibnamefont{Woodley}},
  \bibinfo{journal}{Chem. Mat.} \textbf{\bibinfo{volume}{18}},
  \bibinfo{pages}{1552} (\bibinfo{year}{2006}).

\bibitem[{\citenamefont{Malavasi et~al.}(2002)\citenamefont{Malavasi, Mozzati,
  Azzoni, Chiodelli, and Flor}}]{Malavasi2002}
\bibinfo{author}{\bibfnamefont{L.}~\bibnamefont{Malavasi}},
  \bibinfo{author}{\bibfnamefont{M.~C.} \bibnamefont{Mozzati}},
  \bibinfo{author}{\bibfnamefont{C.~B.} \bibnamefont{Azzoni}},
  \bibinfo{author}{\bibfnamefont{G.}~\bibnamefont{Chiodelli}},
  \bibnamefont{and} \bibinfo{author}{\bibfnamefont{G.}~\bibnamefont{Flor}},
  \bibinfo{journal}{Solid State Communications} \textbf{\bibinfo{volume}{123}},
  \bibinfo{pages}{321} (\bibinfo{year}{2002}).

\bibitem[{\citenamefont{Schiffer et~al.}(1995)\citenamefont{Schiffer, Ramirez,
  Bao, and Cheong}}]{Schiffer1995}
\bibinfo{author}{\bibfnamefont{P.}~\bibnamefont{Schiffer}},
  \bibinfo{author}{\bibfnamefont{A.}~\bibnamefont{Ramirez}},
  \bibinfo{author}{\bibfnamefont{W.}~\bibnamefont{Bao}}, \bibnamefont{and}
  \bibinfo{author}{\bibfnamefont{S.-W.} \bibnamefont{Cheong}},
  \bibinfo{journal}{Phys. Rev. Lett.} \textbf{\bibinfo{volume}{75}},
  \bibinfo{pages}{3336} (\bibinfo{year}{1995}).

\bibitem[{\citenamefont{Yunoki et~al.}(1998)\citenamefont{Yunoki, Hu, Malvezzi,
  Moreo, Furukawa, and Dagotto}}]{Yunoki1998}
\bibinfo{author}{\bibfnamefont{S.}~\bibnamefont{Yunoki}},
  \bibinfo{author}{\bibfnamefont{J.}~\bibnamefont{Hu}},
  \bibinfo{author}{\bibfnamefont{A.~L.} \bibnamefont{Malvezzi}},
  \bibinfo{author}{\bibfnamefont{A.}~\bibnamefont{Moreo}},
  \bibinfo{author}{\bibfnamefont{N.}~\bibnamefont{Furukawa}}, \bibnamefont{and}
  \bibinfo{author}{\bibfnamefont{E.}~\bibnamefont{Dagotto}},
  \bibinfo{journal}{Phys. Rev. Lett.} \textbf{\bibinfo{volume}{80}},
  \bibinfo{pages}{845} (\bibinfo{year}{1998}).

\bibitem[{\citenamefont{Simmons}(1963)}]{Simmons1963}
\bibinfo{author}{\bibfnamefont{J.}~\bibnamefont{Simmons}}, \bibinfo{journal}{J.
  Appl. Phys.} \textbf{\bibinfo{volume}{34}}, \bibinfo{pages}{1793}
  (\bibinfo{year}{1963}).

\bibitem[{\citenamefont{Shang et~al.}(2006)\citenamefont{Shang, Wang, Chen,
  Dong, Li, and Zhang}}]{Shang2006}
\bibinfo{author}{\bibfnamefont{D.}~\bibnamefont{Shang}},
  \bibinfo{author}{\bibfnamefont{Q.}~\bibnamefont{Wang}},
  \bibinfo{author}{\bibfnamefont{L.}~\bibnamefont{Chen}},
  \bibinfo{author}{\bibfnamefont{R.}~\bibnamefont{Dong}},
  \bibinfo{author}{\bibfnamefont{X.}~\bibnamefont{Li}}, \bibnamefont{and}
  \bibinfo{author}{\bibfnamefont{W.}~\bibnamefont{Zhang}},
  \bibinfo{journal}{Phys. Rev. B} \textbf{\bibinfo{volume}{73}},
  \bibinfo{pages}{245427} (\bibinfo{year}{2006}).

\end{thebibliography}

\end{document}